\documentclass[prd,nofootinbib,preprintnumbers,
preprint,
floatfix]{revtex4}
\usepackage{empheq}

\usepackage{amsmath}
\usepackage{amsfonts}
\usepackage{amssymb}
\usepackage{dsfont}
\usepackage{bm}
\usepackage[hyperfootnotes=false]{hyperref}
\usepackage[dvipsnames]{xcolor}
\usepackage{apptools}
\usepackage{enumerate}
\usepackage{subcaption}
\usepackage{diagbox}

\newcommand{\T}{\mathcal T}

\newcommand{\F}{\mathcal F}
\newcommand{\I}{\mathcal I}

\newcommand{\J}{\mathcal J}

\def\theremark{\arabic{section}.\arabic{remark}}
\def\thetheorem{\arabic{section}.\arabic{theorem}}

\def\thedefinition{\arabic{section}.\arabic{definition}}

\renewcommand*{\email}[1]{\footnote{Electronic address: \href{mailto:#1}{\nolinkurl{#1}} }}

\begin{document}

\title{Gravitational shadow and emission spectrum of thin accretion disks in a plasma medium}
\author{Kirill Kobialko${}^{1,\,}$\email{kobyalkokv@yandex.ru}}
\author{Dmitri Gal'tsov${}^{1,\,}$\email{galtsov@phys.msu.ru}}
\author{Alexey Molchanov${}^{1,\,}$\email{alexeybm2009@gmail.com}}
\affiliation{${}^1$ Faculty of Physics, Moscow State University, 119899, Moscow, Russia}

\begin{abstract}
In anticipation of future multi-frequency observations of black hole with the Next Generation Event Horizon Telescope (ngEHT), we construct spectral images of a thin accretion disk around a spherically symmetric black hole immersed in cold, non-magnetized, pressureless plasma. The radiation from the disk is assumed to be thermal, and the surrounding plasma is entrained by its rotation. We use the general relativistic transport equation for the radiation in the plasma, accounting for both dispersion and plasma motion but neglecting absorption. Shadow images and intensity maps are computed across the full spectrum for an inverse power-law plasma density profile. The results show a strong dependence of the observed images on the radiation frequency, which looks promising for the possibility of extracting new information in future observations of the ngEHT.

\end{abstract}

\maketitle

\setcounter{page}{2}

\setcounter{equation}{0}
\setcounter{subsection}{0}

\section{Introduction}

The Event Horizon Telescope (EHT) has produced images of the plasma flows around the supermassive black holes with high enough resolution in a fixed frequency band (around 230 GHz) \cite{EventHorizonTelescope:2019dse,EventHorizonTelescope:2022wkp}. However, black hole images are expected to exhibit complex frequency-dependent structures due to variations in synchrotron emissivity, optical depth, Faraday effects and so on, appealing to multi-frequency studies  (see, e.g., \cite{Moscibrodzka:2017lcu,Chael:2022meh,Ricarte:2022sxg}). The Next Generation Event Horizon Telescope (ngEHT) \cite{Johnson:2023ynn} will extend these capabilities by enabling observations across a broader range of frequencies 86, 230, and 345 GHz, each with a wide bandwidth coverage.
At these frequencies, sources such as Sgr A* and M87* transit  from optically thin to optically thick. Resolved spectral index maps in the near-horizon \cite{Chael:2022meh,Desire:2024mzp} and jet-launching regions \cite{EHT:2025ifc} can clarify properties of the emitting plasma which remain inaccessible in single-frequency observations \cite{Ricarte:2022sxg}. 

On the theoretical side, various analytical and semi-analytical approaches have recently been employed to study the propagation of effective photons in plasma surrounding black holes and their accretion disks \cite{Perlick:2015vta,Perlick:2017fio,Perlick:2023znh,Bezdekova:2022gib,Briozzo:2022mgg,Bogush:2023ojz,Kobialko:2022uzj,Song:2022fdg}. Similar to the vacuum case, these studies have led to consistent constructions of analytical expressions for black hole shadows, whose properties depend on particle energy or radiation frequency \cite{Kobialko:2023qzo,Bezdekova:2022gib}. However, mostly these works do not address the transport of the radiation’s spectral distribution. Additionally, many of the results assume cold, static plasma, although interest in moving plasmas with arbitrary refractive indices is growing \cite{Bezdekova:2024vct}. Numerical ray tracing that incorporates dispersion effects has been explored in several studies \cite{Sareny:2019kxs,Balek:2023rig,Rogers:2016xcc,Rogers:2017ofq,McDonald:2023shx}. Notably, \cite{Rogers:2024lhb} and \cite{Rogers:2015dla} consider ray tracing through absorbing dielectric media in Schwarzschild spacetime and frequency-dependent gravitational lensing in plasma, respectively. Perturbative and other numerical methods have also been discussed in \cite{Crisnejo:2022qlv}. While significant attention has been devoted to both analyzing gravitational shadows in plasma \cite{Kobialko:2023qzo,Atamurotov:2015nra,Abdujabbarov:2015pqp,Abdujabbarov:2016efm,Dastan:2016bfy,Bisnovatyi-Kogan:2017kii,Huang:2018rfn,Babar:2020txt,Chowdhuri:2020ipb,Briozzo:2022mgg,Zhang:2022osx} and modeling emission spectra from thin accretion disks in vacuum \cite{Bogush:2022hop,Gyulchev:2021dvt,Gyulchev:2019tvk,Igata:2025glk,Hou:2022eev}, these approaches have largely progressed independently.

\begin{figure}[tb!]
\centering
\includegraphics[scale=0.6]{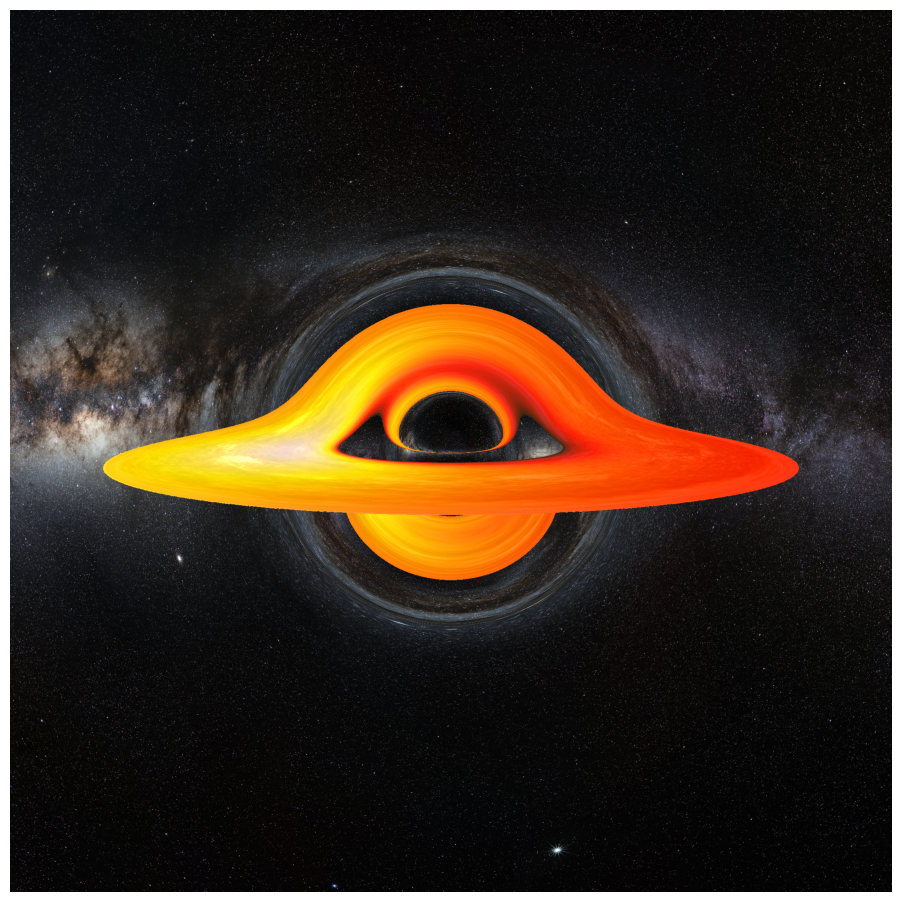} 
\caption{Strong gravitational lensing in a plasma medium.}
\label{SSH_a}
\end{figure}

In this paper, we develop a unified approach combining the Novikov-Thorne thin accretion disk model \cite{Page:1974he,Bambi:2017khi} with analytical and numerical studies of frequency-dependent plasma effects in black hole environments. We examine how cold, non-magnetized, pressureless plasma modifies both the emission spectrum and shadow boundary, considering thermal radiation from the disk propagating through a co-rotating plasma shell \cite{Cole,Gaponenko2023}. The analysis uses the general relativistic radiative transfer equation \cite{Lindquist:1966igj,Kichenassamy:1985zz}, neglecting the absorption and intrinsic radiation of the plasma.

Working in a static, spherically symmetric spacetime, we leverage the symmetry to characterize radiation propagation through equatorial trajectories, with arbitrary emission directions addressed via coordinate rotations. Our framework yields: generalized equations for blackbody radiation transport, and complete spectral maps across the observer's celestial sphere. Representative frequency-resolved images, exemplified in Fig.~\ref{SSH_a}, demonstrate the model's capabilities for probing plasma-mediated astrophysical phenomena.

To compute the full spectrum of observed intensity, we employ custom parallel computation algorithms for numerical ray tracing, complemented by analytical approximations of effective potentials. This approach significantly reduces the computation time for a single spectral slice to well below one second. The numerical scheme is based on the fourth-order Runge–Kutta (RK4) method and is optimized for parallel computations, while consistently taking into account the frequency dependence in radiative transfer.

The article is organized as follows: Section~\ref{sec:radiation_transport} reviews the theoretical framework for thermal radiation and relativistic transport in cold, non-magnetized, pressureless plasma. We derive the observed intensity formula for blackbody radiation in a transparent moving medium, neglecting absorption and self-emission processes. In Section~\ref{sec:spherical_spacetime}, we adapt these formulations to a spherically symmetric spacetime with stationary rotating plasma. Additionally, we verify that condition $\omega\geq\omega_p$ of wave propagation in a moving medium remains valid throughout the domain of motion. Section~\ref{sec:circular_orbits} examines: gravitational shadow boundaries, reflection conditions and emission characteristics of thin accretion disks within the Novikov-Thorne model. Section~\ref{sec:numerical_results} presents a comprehensive analysis of thin accretion disk emission spectra in Schwarzschild spacetime for diverse plasma profiles. Finally, Section~\ref{sec:conclusion} discusses the implications of our results and outlines future research directions.

\section{Relativistic radiation transport in dispersive media}
\label{sec:radiation_transport}

Throughout the paper, we adopt natural units in which \( G = c = \hbar = 1 \), where \( G \) is the Newtonian gravitational constant, \( c \) is the speed of light, and \( \hbar \) is the reduced Planck constant. Additionally, we set the Boltzmann constant \( k_B = 1 \) by defining the temperature \( T \) appropriately.

Recall the main aspects of relativistic radiation transport in dispersive media \cite{Kichenassamy:1985zz}. We are interested in the propagation of radiation in a curved spacetime with the metric \( g_{\alpha\beta} \) through a weakly absorbing, isotropic, and normally dispersive plasma medium, characterized by a 4-velocity \( v^\alpha \) (\( v^\alpha \cdot v_\alpha = -1 \)), and the plasma frequency \cite{Perlick:2015vta}
\begin{equation}
\omega^2_p = \frac{4 \pi e^2}{m_e} N_e(r),
\end{equation}
where \( e \) and \( m_e \) are the charge and mass of an electron, respectively, and \( N_e \) is the electron number density measured in the plasma rest frame. In the geometrical-optics approximation, the propagation of radiation in a cold, non-magnetized, pressureless plasma can be described by the following Hamiltonian \cite{Perlick:2023znh,Kichenassamy:1985zz}
\begin{align} \label{eq:hm}
H=\frac{1}{2}(g^{\alpha\beta} \pi_\alpha \pi_\beta + \omega^2_p),
\end{align}
where $\pi_\alpha$ is the photon 4-momentum and the corresponding Hamilton's equations 
\begin{align} \label{eq:hm_eq}
\dot{\pi}_\alpha =-\partial_\alpha H, \quad \dot{x}^\alpha =\partial_{\pi_\alpha}H= g^{\alpha\beta} \pi_\beta, \quad H=0,
\end{align}
where the dot denotes the derivative with respect to the appropriate affine parameter \( \lambda \). Note that the Hamiltonian does not depend on the plasma velocity, as discussed in Ref. \cite{Bezdekova:2024vct}.

The frequency of radiation measured in the plasma rest frame is given by
\begin{align} \label{eq:loc_frequency}
\omega = - v^\alpha\cdot\pi_\alpha. 
\end{align}
As a consequence, the 4-momentum of an effective photon can be written in the form of a \( 1+3 \) decomposition
\begin{align} 
\pi^\alpha = \omega v^\alpha + k l^\alpha, \quad l^\alpha l_\alpha =1, \quad v^\alpha \cdot l_\alpha=0.
\end{align}
Due to the condition $H=0$ from (\ref{eq:hm}) we get following dispersion relation 
\begin{align} \label{eq:dispersion_relation}
k^2=\omega^2-\omega^2_p.
\end{align}
For the phase and group velocity of radiation flux we have respectively \cite{Rybicki:2004hfl}
\begin{align} \label{eq:ph_g_velocity}
v_{ph}= \frac{\omega}{k}=\frac{1}{\sqrt{1-\omega^2_p/\omega^2}}, \quad v_{g}=\frac{\partial\omega}{\partial k}=\sqrt{1-\omega^2_p/\omega^2},
\end{align}
since in the plasma rest frame, the group velocity does not depend on the direction. In particular, there is a limitation \( \omega \geq \omega_p \) on the frequencies of photons that can propagate in a plasma medium.

One of the main characteristics of the radiation flux is the specific intensity. The specific intensity of the radiation in a plasma medium is given by \cite{Kichenassamy:1985zz}
\begin{align}  \label{eq:intensity_distribution}
I(\omega) = \omega v_{g} f, 
\end{align}
where $f$ is the distribution function, which determines the number of phenomenological photons $dN$ with frequencies in the range $d\omega$ crossing along the rays spatial 3-volume $d\Sigma$ in a solid angle $d\Omega$
\begin{align} \label{eq:number_distribution}
dN = f \cdot d\omega d\Omega d\Sigma.
\end{align}

We will also be interested in the black body radiation spectrum in a plasma medium. Such a spectrum can arise in various models of accretion disks, for example, in the Novikov-Thorne model \cite{Page:1974he,Bambi:2017khi}. We assume that the accretion disk itself is embedded in a transparent plasma medium, which, at least in the small vicinity of the disk, moves with the same velocity. The corresponding distribution function is
\cite{Cole,Gaponenko2023}
\begin{align}  \label{eq:plank_plasma}
f_{plank}(\omega) =\frac{\omega^2}{4\pi^3\cdot v^2_{ph}v_{g}}\cdot \frac{1}{e^{\frac{ \omega}{T}} - 1}.
\end{align}

In order to determine the radiation intensity at the observation point, one should use the relativistic transport equations \cite{Lindquist:1966igj}. However, the commonly used equations are not suitable because they do not take into account the effect of dispersion. Since the derivation of these equations is based on Liouville's theorem, which remains valid for any dynamical system, including non-geodesic motion, there is a generalization for systems with dispersion \cite{Kichenassamy:1985zz}. The flux of photons with momenta in the range \( d\pi = \omega^{-1} k^2 \cdot dk \, d\Omega \) \cite{Lindquist:1966igj,Kichenassamy:1985zz} across a spatial section \( d\Sigma \) orthogonal to the observer's velocity \( v^\alpha \) is given by
\begin{align} \label{eq:invariant_number_distribution}
dN = -\T \cdot v^\alpha \pi_\alpha \cdot d\Sigma \cdot d\pi,
\end{align}
where $\T$ is the invariant relativistic distribution function, which satisfies the following transport equations
\begin{align}
\frac{d\T}{d\lambda} = -\eta \T + \J,
\end{align}
where $\eta$ and $\J$ are, respectively, the invariant absorption and emission coefficients. At the same time, from Eqs. (\ref{eq:number_distribution}), (\ref{eq:invariant_number_distribution}) and  (\ref{eq:loc_frequency}), (\ref{eq:ph_g_velocity}) the connection with the previously introduced distribution function reads as \cite{Kichenassamy:1985zz}
\begin{align} \label{eq:connection_distribution}
\T= k^{-2} \cdot \frac{d\omega}{dk}\cdot f=\omega^{-2}\cdot v^{-1}_{g}\cdot f.
\end{align}
From Eqs. (\ref{eq:intensity_distribution}) and (\ref{eq:connection_distribution}) for specific intensity we find 
\begin{align} \label{eq:T_intensity}
\T = \omega^{-3}\cdot v^{-2}_{g}\cdot I(\omega). 
\end{align}
And finally from Eqs. (\ref{eq:plank_plasma}), (\ref{eq:connection_distribution}) and (\ref{eq:ph_g_velocity}) for black body radiation we get
\begin{align} \label{eq:T_plank}
\T_{plank} = \frac{1}{4\pi^3}\cdot \frac{1}{e^{\frac{ \omega}{T}} - 1},
\end{align}
which does not depend on the plasma frequency at all. Actually, the plasma frequency drop occurred due to relation $v_{ph}\cdot v_{g}=1$ which is universal for dispersion relations of the form (\ref{eq:dispersion_relation}).

Note that the Hamiltonian for cold, non-magnetized, pressureless plasma does not depend on the plasma velocity, as discussed in Refs. \cite{Bezdekova:2024vct,Feleppa:2024vdk}. Therefore, the plasma velocity can be chosen arbitrarily based on reasonable physical considerations. We assume that the plasma undergoes stationary rotation around the same axis as the accretion disk, with some angular velocity \( \Omega_p \). From continuity considerations in the vicinity of the accretion disk, the plasma velocity \( \Omega_p \) coincides with the disk velocity, i.e., \( \Omega_p = \Omega_D \), meaning that the radiation occurs in a plasma at rest. At the observation point, we assume that the plasma becomes static, i.e., \( \bar{\Omega}_p = 0 \). The corresponding 4-velocities are the observer's 4-velocity \( v_o^\alpha = v^\alpha|_o \) and the source's 4-velocity \( v_s^\alpha = v^\alpha|_s \). We assume that the source radiates flux with a Planck spectrum (\ref{eq:plank_plasma}). The observed frequency is related to the source frequency by the usual redshift factor \cite{Bambi:2017khi}.
\begin{align}
\omega_s/\omega_o = 1+z, \quad 1+z=\frac{- v^\alpha_s\cdot\pi_\alpha}{- v^\alpha_o\cdot\pi_\alpha}.
\end{align}
Now we aim to determine the observed specific intensity. In this work, we assume that the plasma medium (excluding the dense accretion disk) is weakly absorbing and does not emit radiation, i.e., \( \eta = 0 \) and \( \mathcal{J} = 0 \). Under these assumptions, the relativistic distribution function \( \mathcal{T} \) remains invariant along the radiation flux. Then, using Eqs.~(\ref{eq:T_intensity}) and (\ref{eq:T_plank}), we obtain
\begin{align} \label{eq:intensity_transport}
I_o(\omega_o)=\frac{\omega^{3}_o}{4\pi^3}\cdot \frac{1-\frac{\bar{\omega}^2_p}{\omega^{2}_o}}{e^{\frac{ (1+z)\omega_o}{T}} - 1}.
\end{align}
Here, the bar indicates that the plasma frequency \( \omega_p^2 \) is evaluated at the observation point. Interestingly, this result does not explicitly depend on the plasma frequency at the emission point. However, since radiation propagates along frequency-dependent trajectories, the observed spectrum at a given point will not follow a pure Planck distribution, but rather represent a superposition of Planck spectra emitted by different regions of the accretion disk.

The applicability of this formula requires that
\begin{align}
\omega_s\geq \omega_p \quad \Rightarrow \quad \omega_o\geq \omega_p/(1+z).
\end{align}
However, as will become clear below, these conditions are automatically satisfied on the equations of motion for the considered stationary plasma rotation. Therefore, the derived equations remain applicable for all observed frequencies (at least within the range of applicability of geometric optics); low-frequency radiation is simply reflected by the plasma. 

\section{Spherically symmetric spacetime}
\label{sec:spherical_spacetime}

Let us consider a general static, spherically symmetric, four-dimensional spacetime with coordinates $x^\alpha = (t, r, \theta, \phi)$ and the following metric tensor:
\begin{equation}  \label{eq:metric_general}
    ds^2 = g_{\alpha\beta} \cdot dx^\alpha dx^\beta= -\alpha dt^2 + \beta  dr^2 + \gamma\left(d\theta^2 + \sin^2\theta d\phi^2\right),
\end{equation}
where the functions $\alpha(r)$, $\beta(r)$, and $\gamma(r)$ depend only on the radial coordinate $r$ and are assumed to be positive, since we consider only the domain of outer communication of the spacetime \cite{Grenzebach:2014fha}. The corresponding Hamiltonian (\ref{eq:hm}) reads as:
\begin{align}
H= -\frac{\pi_t^2}{2\alpha} +\frac{\pi_r^2}{2\beta} +\frac{\pi_\theta^2}{2\gamma}+\frac{\pi_\phi^2}{2\gamma \sin^2\theta} + \frac{1}{2}\omega^2_p.
\end{align}
Since the components of the metric depend only on the coordinates $(r, \theta)$, the Hamiltonian equations (\ref{eq:hm_eq}) for the remaining coordinates $\dot{\pi}_t = 0$ and $\dot{\pi}_\phi = 0$, yield conservation laws for
\begin{align}
\pi_t=-\omega_{\infty}, \quad \pi_\phi=L,
\end{align}
and impact parameter
\begin{align} \label{eq:impact_parameter}
\rho = \omega_{\infty}^{-1}L.
\end{align}
Note that in an asymptotically flat spacetime, in a given parametrization, $\omega_{\infty}$ is the frequency of the radiation measured by a static asymptotic observer $v^\alpha_o = \delta^\alpha_t$, since $-v^\alpha_o \cdot \pi_\alpha = -\pi_t = \omega_{\infty}$ \cite{Perlick:2023znh}.

\begin{figure}[tb!]
\centering
\includegraphics[scale=0.5]{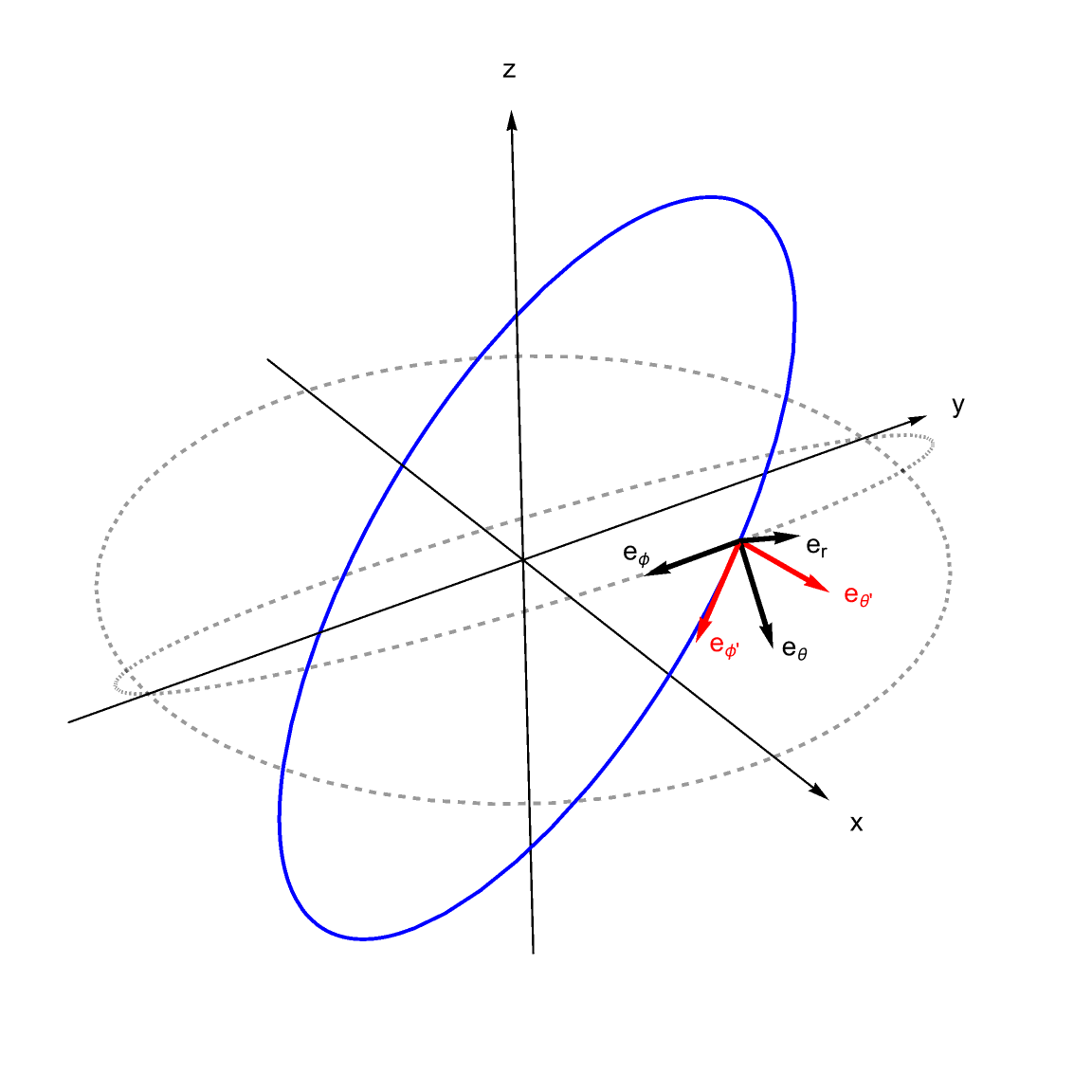} 
\caption{Rotation of the coordinate system and corresponding transformations of the tetrad.}
\label{RA}
\end{figure}

Consider a static observer at point $(\bar{r}, \bar{\theta}, \bar{\phi} = 0)$ with 4-velocity $v_o^\alpha = \bar{\alpha}^{-1/2} \cdot \delta^\alpha_t$ and the radiation with 4-momentum $\pi_\alpha$ arriving to the observer. Here the bar means that the calculations are performed at the observation point, $\bar{\alpha} = \alpha(\bar{r})$. The observed frequency (\ref{eq:loc_frequency}) reads as
\begin{align} \label{eq:observed_frequency}
\omega_{o} =- v_o^\alpha \cdot \pi_\alpha = \bar{\alpha}^{-1/2} \cdot\omega_{\infty}.
\end{align}
Due to the conditions $\omega_{o} =- v_o^\alpha \cdot \pi_\alpha$ and $\pi_\alpha \pi^\alpha=-\bar{\omega}^2_p$ corresponding tangent vector $\dot{x}^\alpha=\pi^\alpha$ can be expanded in an orthonormal tetrad at the observation point as follows 
\begin{align} \label{eq:tetrad_expand}
\dot{x}^\alpha = \omega_{o}\cdot v_o^\alpha + \omega_{o}\cdot\sqrt{1-\bar{\omega}^2_p/\omega_{o}^2}\cdot \left(\cos \Theta e_r{}^\alpha+\sin \Theta \sin \Phi e_\theta{}^\alpha +\sin \Theta \cos \Phi e_\phi{}^\alpha\right), 
\end{align}
where the tetrad $e_i{}^\alpha$ ($i=r,\theta,\phi$) are shown on the Fig. \ref{RA} and reads 
\begin{align}
e_r{}^\alpha = \bar{\beta}^{-1/2}\cdot\delta^\alpha_r, \quad e_\theta{}^\alpha = \bar{\gamma}^{-1/2}\cdot\delta^\alpha_\theta, \quad  e_{\phi}{}^\alpha = -\bar{\gamma}^{-1/2}\sin^{-1}\bar{\theta}\cdot\delta^\alpha_\phi,
\end{align}
where $\Theta$ is the angle between apparent source direction\footnote{The apparent source direction is opposite to the direction of the wave vector (\ref{eq:tetrad_expand})} and the north pole of the celestial sphere, which is directed at the black hole, and $\Phi$ is the second angle on the celestial sphere such that $\Phi = 0$ corresponds to a ray coming counterclockwise relative to the axis of the coordinate system. From Eqs.~(\ref{eq:impact_parameter}), (\ref{eq:observed_frequency}), and (\ref{eq:tetrad_expand}), the impact parameter for such radiation is
\begin{align} \label{eq:impact_parameter_a}
\rho = -\sqrt{\frac{\bar{\gamma}}{\bar{\alpha}}}\cdot\sqrt{1-\bar{\omega}^2_p/\omega_{o}^2}\cdot\sin\bar{\theta} \sin \Theta \cos \Phi.
\end{align}

The spherical symmetry of the system implies that all trajectories lie in planes containing the coordinate origin. While the equations of motion take their simplest form in the equatorial plane ($\theta = \pi/2$), we must account for non-equatorial trajectories. To achieve this, we employ the coordinate rotations illustrated in Fig.~\ref{RA}, which enable complete characterization of general orbital motion while maintaining computational efficiency. Let us define the Cartesian coordinates corresponding to $(\theta, \phi)$ as
\begin{align}
x= \cos \phi \sin \theta, \quad
y= \sin \phi \sin \theta, \quad
z = \cos \theta.
\end{align}
Now we will perform two successive rotation coordinate transformations
\begin{align}
\left(\begin{array}{c}
 x' \\
 y'  \\
 z' \\
\end{array}\right)= O_\Phi O_\theta \left(\begin{array}{c}
 x\\
 y  \\
 z \\
\end{array}\right),
\end{align}
where
\begin{align}
O_\theta=\left(
\begin{array}{ccc}
 \sin \bar{\theta} & 0 & \cos \bar{\theta} \\
 0 & 1 & 0 \\
 -\cos \bar{\theta} & 0 & \sin \bar{\theta} \\
\end{array}
\right), \quad O_\Phi=\left(
\begin{array}{ccc}
 1 & 0 & 0 \\
 0 & \cos \Phi & \sin \Phi \\
 0 & -\sin \Phi & \cos \Phi \\
\end{array}
\right),
\end{align}
and also introduce the coordinates $ (\theta',\phi')$
\begin{align}
x'= \cos \phi' \sin \theta', \quad
y'= \sin \phi' \sin \theta', \quad
z' = \cos \theta'.
\end{align}
Since the spherically symmetric metric is invariant under rotations ($x^2 + y^2 + z^2 = x'^2 + y'^2 + z'^2$), in the new coordinates $x^{\alpha'} = (t, r, \theta', \phi')$ the metric has the same form as (\ref{eq:metric_general}).
\begin{equation} 
    ds^2 = -\alpha dt^2 + \beta  dr^2 + \gamma\left(d\theta'^2 + \sin^2\theta' d\phi'^2\right).
\end{equation}
The meaning of these rotations is clearly seen in Fig. \ref{RA}. Rotation $O_\theta$ means that in the new coordinates $(\theta', \phi')$ the observer $(\theta = \bar{\theta}, \bar{\phi} = 0)$ is located at the point $(\theta' = \pi/2, \phi' = 0)$, i.e., in the equatorial plane. While the second rotation $O_\Phi$ rotates the tetrad in such a way that the vector $\pi^{\alpha'}$ has zero projection onto $e_{\theta'}{}^{\alpha'} = \bar{\gamma}^{-1/2} \cdot \delta^{\alpha'}_{\theta'}$, i.e., the trajectory lies entirely in the equatorial plane $\theta' = \pi/2$. In particular, we have (the angle $\Theta$ is not transformed because rotations do not affect the sector $(t,r)$).
\begin{align} \label{eq:tetrad_expand_p}
\dot{x}^{\alpha'}  =\omega_{o} \cdot v_o^{\alpha'} +\omega_{o} \cdot \sqrt{1-\bar{\omega}^2_p/\omega_{o}^2}\cdot \left(\cos \Theta e_r{}^{\alpha'} +\sin \Theta e_{\phi'}{}^{\alpha'}\right),
\end{align}
where
\begin{align}
e_r{}^{\alpha'}  = \bar{\beta}^{-1/2}\cdot\delta^{\alpha'}_r, \quad e_{\phi'}{}^{\alpha'} = -\bar{\gamma}^{-1/2}\cdot\delta^{\alpha'}_{\phi'}.
\end{align}
For the inverse transformation generated by the matrix $O^{-1}_\theta O^{-1}_\Phi$ we find the following coordinate transformation law:
\begin{align}  \label{eq:coordinate_rot}      
&\theta = \arccos\left(\cos\phi' \cos\bar{\theta}+ \sin\phi' \sin\Phi \sin\bar{\theta}\right), \\
&\phi = \arctan\left(\frac{ \sin\phi' \cos\Phi}{\cos\phi' \sin \bar{\theta}   -\sin\phi' \sin\Phi \cos \bar{\theta}}\right).  
\end{align} 
Since in coordinates $x^{\alpha'} = (t, r, \theta', \phi')$ the motion occurs in the equatorial plane $\theta'=\pi/2$, we can define the following effective three-dimensional metric:
\begin{equation} \label{eq:metric_3d}
    ds^2_{3D} = -\alpha dt^2 + \beta  dr^2 + \gamma d\phi'^2. 
\end{equation} 
The corresponding Hamiltonian (\ref{eq:hm}) reads as
\begin{align}
H= -\frac{1}{2\alpha}\pi_t^2 +\frac{1}{2\beta}\pi_r^2 +\frac{1}{2\gamma}\pi_{\phi'}^2 + \frac{1}{2}\omega^2_p.
\end{align}
Since the components of the metric depend only on coordinate $r$, the Hamiltonian equations for the remaining coordinates $\dot{\pi_t}=0$ and  $\dot{\pi}_{\phi'}=0$ give conservation laws for
\begin{align}
\pi_t=-\omega_{\infty}, \quad \pi_{\phi'}=L'.
\end{align}
The corresponding impact parameter 
\begin{align} \label{eq:impact_parameter_new}
\rho'=\omega^{-1}_{\infty}L'= -\sqrt{\frac{\bar{\gamma}}{\bar{\alpha}}}\cdot\sqrt{1-\bar{\omega}^2_p/\omega_{o}^2}\sin \Theta =\sin^{-1}\bar{\theta} \cos^{-1} \Phi \cdot \rho.
\end{align}
In order to eliminate the remaining  quadratic term $\pi^2_r$ in Hamilton's equations we introduce a new radial impulse
\begin{align}
p = \omega_{\infty}^{-1}\pi_r/\beta  \quad \Rightarrow \quad \omega_{\infty}^{-1} \dot{r}= p.
\end{align}
Then on the equations of motion (\ref{eq:hm_eq}) 
\begin{align}
\omega_{\infty}^{-1} \dot{p}=-\beta^{-1}\partial_r H \omega_{\infty}^{-2}   + \pi^2_r\beta^{-1}\cdot\partial_r\beta^{-1}\omega_{\infty}^{-2} = -\partial_r \left(\beta^{-1}\left(H-\frac{1}{2\beta}\pi_r^2\right)\right)\omega_{\infty}^{-2}, 
\end{align}
and the complete system of Hamilton's equations (\ref{eq:hm_eq}) reads as:
\begin{align}
\omega_{\infty}^{-1}\cdot\dot{p} = \frac{1}{2} \cdot \partial_r V, \quad \omega_{\infty}^{-1}\cdot\dot{r}=p, \quad \omega_{\infty}^{-1}\cdot\dot{\phi'} =\gamma^{-1}\rho', \quad \omega_{\infty}^{-1}\cdot\dot{t} =\alpha^{-1}, \quad p^2=V, 
\end{align}
where effective potential 
\begin{align} \label{eq:effective_potential}
V= \beta^{-1}\left(\alpha^{-1}-\gamma^{-1}\rho'^2-\omega^2_p / \omega_{\infty}^{2}\right).
\end{align}
Using partition (\ref{eq:tetrad_expand_p}) we can formulate the following initial data for the system of Hamilton's equations
\begin{align} \label{eq:initial_data}
p=\left(\bar{\alpha}\bar{\beta}\right)^{-1/2}\sqrt{1-\bar{\omega}^2_p/\omega_{o}^2} \cdot \cos \Theta, \quad \rho'= -\left(\bar{\gamma}/\bar{\alpha}\right)^{1/2} \sqrt{1-\bar{\omega}^2_p/\omega_{o}^2} \cdot \sin \Theta.
\end{align}
Now we change the parameterization $\lambda\rightarrow \lambda'=\omega_\infty \cdot \lambda$ which leaves the trajectories invariant and unifies the step of numerical integration for different $\omega_\infty$. We get
\begin{align} \label{eq:equations_numerical}
\dot{p} = \frac{1}{2} \cdot \partial_r V, \quad \dot{r}=p, \quad \dot{\phi'} =\gamma^{-1}\rho', \quad \dot{t} =\alpha^{-1}, \quad p^2=V.
\end{align}
In our numerical analysis, we employ a fourth-order Runge-Kutta (RK4) integration scheme with negative step size to solve the equations backward in time. At each integration step, we apply the coordinate transformation (\ref{eq:coordinate_rot}) to reconstruct photon paths in the original coordinate system.

But first, we define more convenient coordinates for the celestial sphere using a stereographic projection.
\begin{align}
X = 2 \tan\left(\Theta/2\right) \cos \Phi, \quad Y = 2 \tan\left(\Theta/2\right) \sin \Phi,
\end{align}
or for inverse transformation
\begin{align}
\Theta = 2 \arctan\left(\frac{\sqrt{X^2+Y^2}}{2}\right), \quad \Phi=\arctan(X,Y).
\end{align}
In this case
\begin{align}
\sin \Phi &= \frac{Y}{\sqrt{X^2+Y^2}}, \quad \cos\Phi = \frac{X}{\sqrt{X^2+Y^2}}, \\
\sin \Theta &= \frac{4\sqrt{X^2+Y^2}}{X^2+Y^2+4}, \quad \cos \Theta = \frac{4-X^2-Y^2}{X^2+Y^2+4}.
\end{align}
These coordinates will serve as our primary framework for subsequent analysis.

We must also verify the model's validity by ensuring the frequency condition $\omega \geq \omega_p$ holds throughout our analysis. For this purpose, we first define the 4-velocity of a stationary, uniformly rotating plasma:
\begin{align}
v^\alpha =\frac{\delta^\alpha_t+\Omega_p\delta^\alpha_\phi}{\sqrt{\alpha-\gamma\Omega^2_p}},
\end{align}
where $\Omega_p$ is the angular velocity of plasma rotation in the original coordinates $x^\alpha = (t, r, \theta, \phi)$. The redshift factor relating the emitted frequency $\omega$ to the observed frequency $\omega_o$ then takes the form
\begin{align} \label{eq:redshift_1}
1+z=\frac{-v^\alpha \pi_\alpha}{-v^\alpha_o \pi_\alpha}=\frac{\omega_\infty-\Omega_p L}{\omega_{o}\sqrt{\alpha-\gamma\Omega^2_p}}=\left(1-\Omega_p \rho\right)\cdot \sqrt{\frac{\bar{\alpha}}{\alpha-\gamma\Omega^2_p}},
\end{align}
where $\pi_\alpha$ and $\rho$ are photon 4-momentum and impact parameter in the original coordinate system. In particular
\begin{align} \label{eq:redshift_2}
\omega = \frac{1-\Omega_p \rho}{\sqrt{\alpha-\gamma\Omega^2_p}} \cdot \omega_{\infty}.
\end{align}
Let us now recall that motion is possible only in the region where the effective potential $V$ (\ref{eq:effective_potential}) is non-negatively defined, i.e
\begin{align}
\alpha^{-1}-\gamma^{-1}\rho'^2-\omega^2_p / \omega_{\infty}^{2}\geq0 \quad \Rightarrow \quad \alpha^{-1}-\gamma^{-1}\rho^2-\omega^2_p / \omega_{\infty}^{2}\geq0,
\end{align}
since $\rho'\geq\rho$ by virtue of Eq. (\ref{eq:impact_parameter_new}). Then we have
\begin{align}
\omega^2_p / \omega_{\infty}^{2}\leq\alpha^{-1}-\gamma^{-1}\rho^2.
\end{align}
From Eq. (\ref{eq:redshift_2}) it then follows that
\begin{align} \label{eq:applicability}
\omega^2_p / \omega^{2}&\leq\frac{\left(1-\alpha\gamma^{-1}\rho^2\right)\left(1-\alpha^{-1}\gamma\Omega_p^2\right)}{\left(1-\Omega_p\rho\right)^2}\leq1-\frac{\alpha}{\gamma}\frac{(\rho-\gamma\Omega_p/\alpha)^2}{\left(1-\Omega_p\rho\right)^2}\leq1.
\end{align}
Thus, the condition $\omega \geq \omega_p$ for wave propagation in a moving medium is always fulfilled and there is no need to check it additionally because geodesics cannot reach forbidden area. Note that strict equality $\omega = \omega_o$ can only arise under the condition that this is a turning point — which is consistent with the vanishing of the group velocity.

\section{Circular orbits, shadow and thin accretion disk} 
\label{sec:circular_orbits}

Circular orbits are fundamental for characterizing both gravitational shadows and the Novikov-Thorne accretion disk model. Circular orbits are determined by the conditions $p=0$ and $\dot{p}=0$ or by virtue of the Eqs. (\ref{eq:equations_numerical}) $V=0$ and $\partial_r V=0$, i.e.
\begin{align} \label{eq:circular_orbits_eq}
\alpha^{-1}-\gamma^{-1}\rho'^2-\omega^2_p/\omega^{2}_\infty=0,\quad
-\alpha^{-2} \partial_r\alpha+\rho'^2 \gamma^{-2}\partial_r\gamma-\omega^{-2}_\infty \partial_r\omega^2_p=0.
\end{align}
Solving this system of equations we find
\begin{align} \label{eq:circular_orbits_sol}
\omega^{2}_\infty=\alpha^2\cdot\frac{\partial_r(\gamma \cdot \omega^2_p)}{\alpha\partial_r\gamma-\gamma\partial_r\alpha}, \quad \rho'^{2}=\frac{\gamma^2}{\alpha^2}\cdot\frac{\partial_r(\alpha \cdot \omega^2_p)}{\partial_r(\gamma \cdot \omega^2_p)}, \quad L'^2=\gamma^2\cdot\frac{\partial_r(\alpha \cdot \omega^2_p)}{\alpha\partial_r\gamma-\gamma\partial_r\alpha}.
\end{align}
This leads directly to the expression for the angular velocity of rotation, which can be derived as follows:
\begin{align}
\Omega'=\frac{d\phi'}{dt}=\frac{\alpha \rho'}{\gamma}=\pm\sqrt{\frac{\partial_r(\alpha \cdot \omega^2_p)}{\partial_r(\gamma \cdot \omega^2_p)}}.
\end{align}
It is well known that the boundary of the gravitational shadow is formed by geodesics asymptotically winding onto such circular orbits or, more generally, a photon surface \cite{Virbhadra:1999nm,Claudel:2000yi,Virbhadra:2002ju,Virbhadra:2022ybp}. In particular, the impact parameter (\ref{eq:impact_parameter_new})  for geodesics forming a shadow must coincide with the impact parameter of circular orbits. Rather than employing the precomputed solution for $\rho'$ from (\ref{eq:circular_orbits_sol}), we solve the first equation of system (\ref{eq:circular_orbits_eq}) directly to determine:
\begin{align}
 \sin^2 \Theta=\frac{\gamma}{\alpha}\cdot \frac{\bar{\alpha}}{\bar{\gamma}}\cdot\frac{1 -\alpha\bar{\alpha}^{-1}\cdot\omega^2_p/\omega^{2}_o}{1-\bar{\omega}^2_p/\omega_{o}^2}, \quad \omega^{2}_o=\bar{\alpha}^{-1}\alpha^2\cdot\frac{\partial_r(\gamma \cdot \omega^2_p)}{\alpha\partial_r\gamma-\gamma\partial_r\alpha}.
\end{align}
The expression for the shadow can be obtained directly from (\ref{eq:circular_orbits_sol}), but the resulting equivalent formula turns out to be less transparent for comparing shadows with the vacuum case and, in particular, for constructing perturbation theory \cite{Kobialko:2024zhc}. In addition, this formula does not depend on the angle $\Phi$ which means that the shadow is just a circle with center at the north pole of the celestial sphere.  Note that, this formula differs from Ref. \cite{Perlick:2015vta,Kobialko:2024zhc}, since it uses not the asymptotic frequency $\omega_{\infty}$ but the locally observed $\omega_{o}$ one. 

It is interesting that the shadow is completely absent if for some $r_e>r_h$ (where $r_h$ the coordinate of the event horizon) there exists a solution of equations
\begin{align} \label{eq:equilibrium_r}
\partial_r(\alpha\omega^2_p)=0,
\end{align}
which by virtue of (\ref{eq:circular_orbits_sol}) and (\ref{eq:impact_parameter_new}) means that $\rho'=0$ and $\Theta=0$ respectively. Corresponding frequency reads as
\begin{align} \label{eq:equilibrium_omega}
\omega^{2}_e =\alpha\bar{\alpha}^{-1}\cdot\omega^2_p. 
\end{align}
The physical meaning of this can be understood by analyzing the effective potential (\ref{eq:effective_potential}). The point $r_e$ for which the shadow disappears is a circular orbit for geodesics with $\rho'=0$. Therefore, in fact, this is not a circular orbit but rather a point of radial equilibrium between gravity and the plasma repulsive force. Moreover, for such frequencies all non-radial geodesics are reflected. Indeed, we have $V(r_e,\rho'=0)=0$. But then for any $\rho'>0$ there will be another turning point at some $r>r_e$ since $V(r_e,\rho'>0)<0$ and at the observation point we assume that $V(\bar{r},\rho'>0)>0$. Thus, the existence of a radial equilibrium point means complete reflection of light at sufficiently low frequencies (see \cite{Rogers:2016xcc} for discussion). It should be noted that such strong gravitational lensing effects can arise also in a vacuum for weakly naked singularities \cite{Virbhadra:2002ju}.

For an asymptotic observer ($\bar{\alpha}=1$, $\bar{\gamma}=\bar{r}^2$ and $\bar{\omega}^2_p=0$)
\begin{align} \label{eq:shadow_asymptotic}
R^2= \frac{\gamma}{\alpha}\cdot \left(1 -\alpha\cdot\omega^2_p/\omega^{2}_o\right), \quad \omega^{2}_o=\alpha^2\cdot\frac{\partial_r(\gamma \cdot \omega^2_p)}{\alpha\partial_r\gamma-\gamma\partial_r\alpha},
\end{align}
where $R^2=\lim_{\bar{r}\rightarrow\infty}R^2_{proj}\cdot\bar{r}^2$ and $R^2_{proj}=X^2+Y^2$.

In this work, we extend beyond the well-studied gravitational shadow phenomenon \cite{Grenzebach:2014fha,Grenzebach:2015oea} to analyze radiation transfer from accretion disks \cite{Bogush:2022hop,Gyulchev:2021dvt,Gyulchev:2019tvk}. For this purpose, we adopt the Novikov-Thorne model \cite{Page:1974he,Bambi:2017khi} of thin accretion disks, which provides a rigorous relativistic framework for describing disk emission properties. According to this model, the disk is located in the equatorial plane $\theta=\pi/2$ of the original coordinate system ($\phi'=\phi$ and $\theta'=\theta$) and performs almost circular geodesic motion. Choosing a constant $\omega^2_p=1$ and defining $\omega_\infty=E_D$, $L'=L_D$, we find
\begin{align}
E_D=\frac{\alpha}{\sqrt{\alpha-\gamma\Omega^2_D}}, \quad L_D=\frac{\gamma \Omega}{\sqrt{\alpha-\gamma\Omega^2_D}}, \quad \Omega_D=\pm\sqrt{\partial_r\alpha/\partial_r\gamma}.
\end{align}
The disc 4-velocity 
\begin{align}
v^\alpha_D = \frac{\delta^\alpha_t+\Omega_D\delta^\alpha_\phi}{\sqrt{\alpha-\gamma\Omega^2_D}}.
\end{align}
In accordance with our model and the conditions of continuous plasma rotation, we assume that at the disk points $\Omega_p=\Omega_D$. In this case, from Eq. (\ref{eq:redshift_1}) for the redshift we immediately find 
\begin{align} \label{eq:red_m}
1+z=\left(1-\Omega_D \rho\right)\cdot \sqrt{\frac{\bar{\alpha}}{\alpha-\gamma\Omega^2_D}}.
\end{align}
Here $\rho\neq\rho_D=L_D/E_D$, because the circular orbit differs from the observed geodesic and is determined at the observation point by Eq. (\ref{eq:impact_parameter_a}).

In Novikov-Thorne model, an expression for the time-averaged energy flux from the surface of the disk in a vacuum as a function of the radial coordinate $r$ reads as 
\begin{align} \label{eq:flux}
\F_{D} = - \frac{\dot{M}}{4 \pi \sqrt{-\det g_{3D}}}\cdot\frac{ \partial_r \Omega_D}{(E_D-\Omega_D L_D)^2} \int^r_{r_{ISCO}} (E_D-\Omega_D L_D)\cdot \partial_r L_D \cdot dr,
\end{align}
where $\det g_{3D}=-\alpha\beta\gamma$ determinant of the metric induced on the equatorial plane (more precisely, the equatorial cylinder $\theta=\pi/2$), $r_{ISCO}$ is an innermost stable circular orbit (ISCO) determined by equations $V=\partial_rV=\partial^2_rV=0$ and $\dot{M}$ is a mass accretion rate. This flux corresponds to the total radiation of a black body in a vacuum with temperature $T_D$ according to the Stefan-Boltzmann law 
\begin{align}
\F_{D} = \sigma T^4_D, \quad \sigma = \frac{\pi^2}{60}.
\end{align}
The vacuum specific intensity of the radiation emitted by the disk is \cite{Bambi:2017khi}
\begin{align}
I_D(\omega_D)= \frac{\omega^3_D}{4\pi^3}\cdot \frac{1}{e^{\frac{ \omega_D}{T_D}} - 1}.
\end{align}

Now we need to compare the characteristics of the radiation of the accretion disk in a vacuum and in the medium. Our assumption is that in a plasma medium the radiation occurs at the same temperature $T=T_D$ but according to the modified Planck law (\ref{eq:plank_plasma}). In this case, as shown in the previous section, the observed intensity will be given by formula (\ref{eq:intensity_transport}) and read as   
\begin{align} \label{eq:intensity_o}
I_{o}(\omega_{o})= \frac{ \omega^3_{o}}{4\pi^3}\cdot \frac{1-\frac{\bar{\omega}^2_p}{\omega^2_{o}}}{e^{\frac{(1+z)\omega_{o}}{T_D(\omega_{o})}} - 1}, \quad T_D(\omega_{o})=\sigma^{-1/4} \F^{1/4}_{D}(\omega_{o}),
\end{align}
where $\F_D(\omega_{o})$ determined by ray tracing and Eq. (\ref{eq:flux}). If the radiation trajectory does not pass through the disk, we will simply assign $T(\omega_{o})$ the temperature of the distant background (microwave background radiation) or near horizon sphere (Hawking radiation). Total received intensity of radiation reads as 
\begin{align} \label{eq:sum}
\I = \int^\infty_0 I_{o}(\omega_{o}) d\omega_{o}. 
\end{align}

One of the important free parameters of the model is mass accretion rate $\dot{M}$ which does not depend on the radial coordinate and can vary over a wide range. Changing this parameter shifts the effective radiation temperature and, in particular, the position of the spectrum maximum by a constant the same for the entire disk. However, in the absence of plasma, normalized total intensity $\I$ distribution remains invariant. Indeed, if there is no plasma total observed intensity easily integrated by changing the variable since neither temperature nor redshift depends on frequency
\begin{align}
\I = \frac{1}{\pi}\cdot\frac{\F_D}{(1+z)^4}. 
\end{align}
When the accretion rate changes, the total intensity will simply receive a new constant multiplier that is the same at all points on the disk. Therefore, normalized expressions such as $\I/\I_{max}$ do not depend on $\dot{M}$. In the case of frequency-dependent radiation, due to dispersion, this parameter has a great influence on total intensity, since it will encode the volume of contributions at different frequencies. 

Another important characteristic of the radiation of the accretion disk is the total observable radiation flux \cite{Bambi:2017khi}
\begin{align} \label{eq:flux_total}
F(\omega_o) = \int I_o(\omega_o) d\Omega, \quad d\Omega = \sin \Theta d\Theta  d\Phi= \frac{16\cdot dX dY}{(4 +X^2+Y^2)^2},
\end{align}
where $d\Omega$ is the element of the solid angle. Being frequency-dependent scalar quantities, these values are most effectively visualized through graphical representations. For clarity, we normalize all such characteristics to their corresponding vacuum values.

\section{Analytical and numerical results} 
\label{sec:numerical_results}

In this section we analyze in detail the frequency-dependent effects of the thin accretion disk radiation. As our canonical example, we employ the Schwarzschild metric:
\begin{equation}  \label{eq:metric_schwarzschild}
    ds^2 = -\left(1-\frac{2M}{r}\right) dt^2 + \left(1-\frac{2M}{r}\right)^{-1}  dr^2 + r^2\left(d\theta^2 + \sin^2\theta d\phi^2\right), 
\end{equation}
and plasma profiles
\begin{equation} 
 \omega^2_p=\frac{M^\sigma}{ r^{\sigma}}, \quad \sigma\geq0.
\end{equation}
In general, this distribution can be multiplied by any constant factor $h$, but this will only lead to frequency renormalization and will not affect the motion pattern. Moreover, in the spectrum, by simultaneously redefining the mass accretion rate, this influence can also be eliminated with an accuracy of some normalization. Indeed, let's say at frequency $\omega_o$ we found redshift factor $1+z$ and temperature $T_D$ of the source using ray tracing. Let's define a new plasma frequency $\omega'^2_p=h^2 \omega^2_p$. Then at frequency $\omega'_o=h\omega'_o$ we have the same $z'=z$ and $T'_D=T_D$ since the effective potential (\ref{eq:effective_potential}) and initial conditions (\ref{eq:initial_data}) includes only the ratio $\omega^2_p/\omega^2_\infty$. If we simultaneously rescale the mass accretion rate such that the modified flux $\mathcal{F}'_D$ satisfies:
\begin{equation}
\mathcal{F}'_D = h^4 \mathcal{F}_D,
\end{equation}
where $\mathcal{F}_D$ is the original disk flux, we obtain:
\begin{align}
I'_{o}(\omega'_{o})= \frac{ \omega'^3_{o}}{4\pi^3}\cdot \frac{1-\frac{\bar{\omega'}^2_p}{\omega'^2_{o}}}{e^{\frac{(1+z')\omega'_{o}}{T'_D(\omega'_{o})}} - 1}=h^3\cdot\frac{ \omega^3_{o}}{4\pi^3}\cdot \frac{1-\frac{\bar{\omega}^2_p}{\omega^2_{o}}}{e^{\frac{(1+z)\omega_{o}}{T_D(\omega_{o})}} - 1}=h^3\cdot I_{o}(\omega_{o}).
\end{align}
Thus, it makes sense to consider only plasma frequencies of type $\frac{M^\sigma}{ r^{\sigma}}$ for different mass accretion rate while everything else can be obtained analytically by simply multiplying by the factor $h$ to the appropriate power.

\begin{figure}[tb!]
\centering
\includegraphics[scale=0.7]{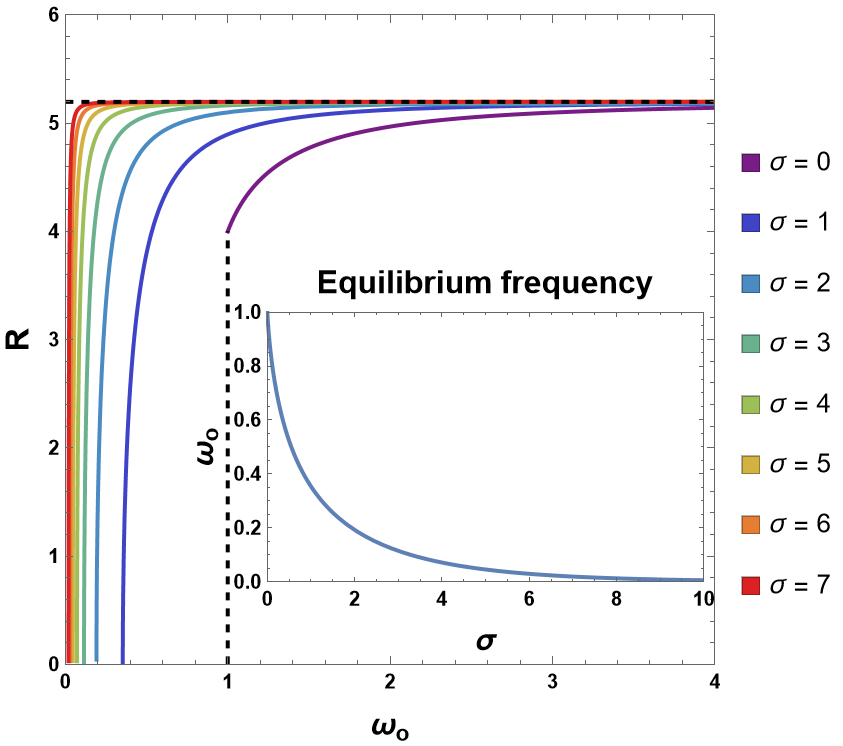} 
\caption{Gravitational shadow radius $R$ as a function of observed frequency $\omega_o$ for a family of plasma frequencies $\omega^2_p=M^\sigma \cdot r^{-\sigma}$ and  equilibrium frequency in Schwarzschild metric.}
\label{SSH}
\end{figure}

For the asymptotic gravitational shadow radius $R$ and observed frequency $\omega_o$ derived from Eq. (\ref{eq:shadow_asymptotic}), we obtain:
\begin{equation} \label{eq:shadow_sch}
 \omega^2_o=\frac{(\sigma -2) (r-2 M)^2 M^{\sigma } r^{-\sigma -1}}{2 (3 M-r)}, \quad R^2=\frac{r^3 (r \sigma -2 M (\sigma +1))}{(\sigma -2) (r-2 M)^2}.
\end{equation}
Given the non-negativity conditions $\omega_o^2 \geq 0$ and $R^2 \geq 0$, the radial coordinate $r$ must satisfy:
\begin{align}
&\sigma = 0, \quad 2M \leq r \leq 4M, \\
0 < &\sigma <2, \quad  3M \leq r \leq 2 M\cdot\frac{\sigma+1}{\sigma},\\
&\sigma = 2, \quad r=3M, \quad R^2= M^2(27-\omega^{-2}_o), \\
2<&\sigma<\infty , \quad  2 M\cdot\frac{\sigma+1}{\sigma} \leq r \leq 3M.
\end{align}
The $\sigma=2$ plasma profile ($\omega_p^2 \propto r^{-2}$) is particularly significant. Although the formal expressions (\ref{eq:shadow_sch}) for the shadow have a singularity if we look at the intervals of $ 3M \leq r \leq 2 M\cdot\frac{\sigma+1}{\sigma}$ for $\sigma <2$ and $ 2 M\cdot\frac{\sigma+1}{\sigma} \leq r \leq 3M$ for $\sigma >2$ and noticing that $\frac{\sigma+1}{\sigma}=3/2$ for $\sigma=2$, it is clear, that we are dealing with indeterminate form $0/0$. To regularize this behavior, we consider a family of frequencies $\omega_s$ and shadows $R_s$ parameterized by the parameter $s$ 
\begin{align}
r_s = 3 M s + 2(1-s) M\cdot\frac{\sigma+1}{\sigma}, \quad 0\leq s\leq1.
\end{align}
Substituting these relations into the shadow equations (\ref{eq:shadow_sch}), we find that for this parameter family, the limit $\sigma \to 2$ is well-defined and yields:
\begin{align}
R^2_s=27 M^2 s, \quad \omega^2_s=\frac{1}{27 (1-s)}.
\end{align}
By combining these equations and omitting the index $s$ we find the exact analytical expression for $\sigma=2$
\begin{align} \label{eq:sigma_2}
R^2= M^2(27-\omega^{-2}_o).
\end{align}
Note that, for $\sigma=0$ the expression for $R$ can be found in the Ref. \cite{Kobialko:2024zhc}. Intermediate cases are shown in the Fig. \ref{SSH}. Note that for a homogeneous plasma $\sigma=0$, since it does not vanish at infinity, there is a lower limit on the frequencies $\omega_o$ and the shadow does not disappear completely. 

The equilibrium radius $r_e$, where gravitational attraction balances the plasma's reflective force, is obtained by solving Eq. (\ref{eq:equilibrium_r}) (the same for both the asymptotic and arbitrary static observer)
\begin{align}
r_e=2M \cdot \frac{\sigma+1}{\sigma}.
\end{align}
The corresponding frequency (\ref{eq:equilibrium_omega}) for a for an arbitrary static observer \cite{Rogers:2016xcc}
\begin{align} \label{eq:equilibrium_f}
\omega^2_e=\frac{\left(2+\frac{2}{\sigma }\right)^{-\sigma }}{1+\sigma}\cdot\frac{1}{1-2 M/\bar{r}}.
\end{align}
At frequencies $\omega_o\leq\omega_e$ the geodesics are completely reflected and the observer observes only the outer region of the universe but not the horizon - the shadow is completely absent. We present the corresponding frequency in Fig. \ref{SSH} for an asymptotic observer. 

We now present our numerical framework for modeling the accretion disk's radiation transfer. The observer is located at a point with coordinates $(\bar{r}=40M, \bar{\theta})$. For a given observed frequency $\omega_o$ at each vertex $(X_{ij},Y_{ij})$ of a rectangular grid on the celestial sphere stereographic projection, we solve the system (\ref{eq:equations_numerical}) using a parallelized fourth-order Runge-Kutta integration scheme with negative step. At each integration step, we apply the coordinate rotations specified in (\ref{eq:coordinate_rot}). The integration terminates when any of the following conditions is satisfied:
\begin{itemize}
    \item The solution approaches within a small neighborhood of the event horizon at $r_h = 2M$
    \item The solution reaches the distant background sphere at $r = 50M$
    \item The solution intersects the accretion disk in the region $6M < r_D < 20M$ with $\theta = \pi/2$ (note that for this metric, ISCO occurs at $r = 6M$)
\end{itemize}

From the obtained solutions, we compute:
\begin{itemize}
    \item The redshift $z$ using Eq. (\ref{eq:red_m})
    \item The temperature $T$ via Eq. (\ref{eq:flux})
    \item The specific intensity via Eq. (\ref{eq:intensity_o}).
\end{itemize}

\begin{figure}[tb!]
\centering
\subfloat[][$\bar{\theta}=84^\circ$]{
  \includegraphics[scale=0.068]{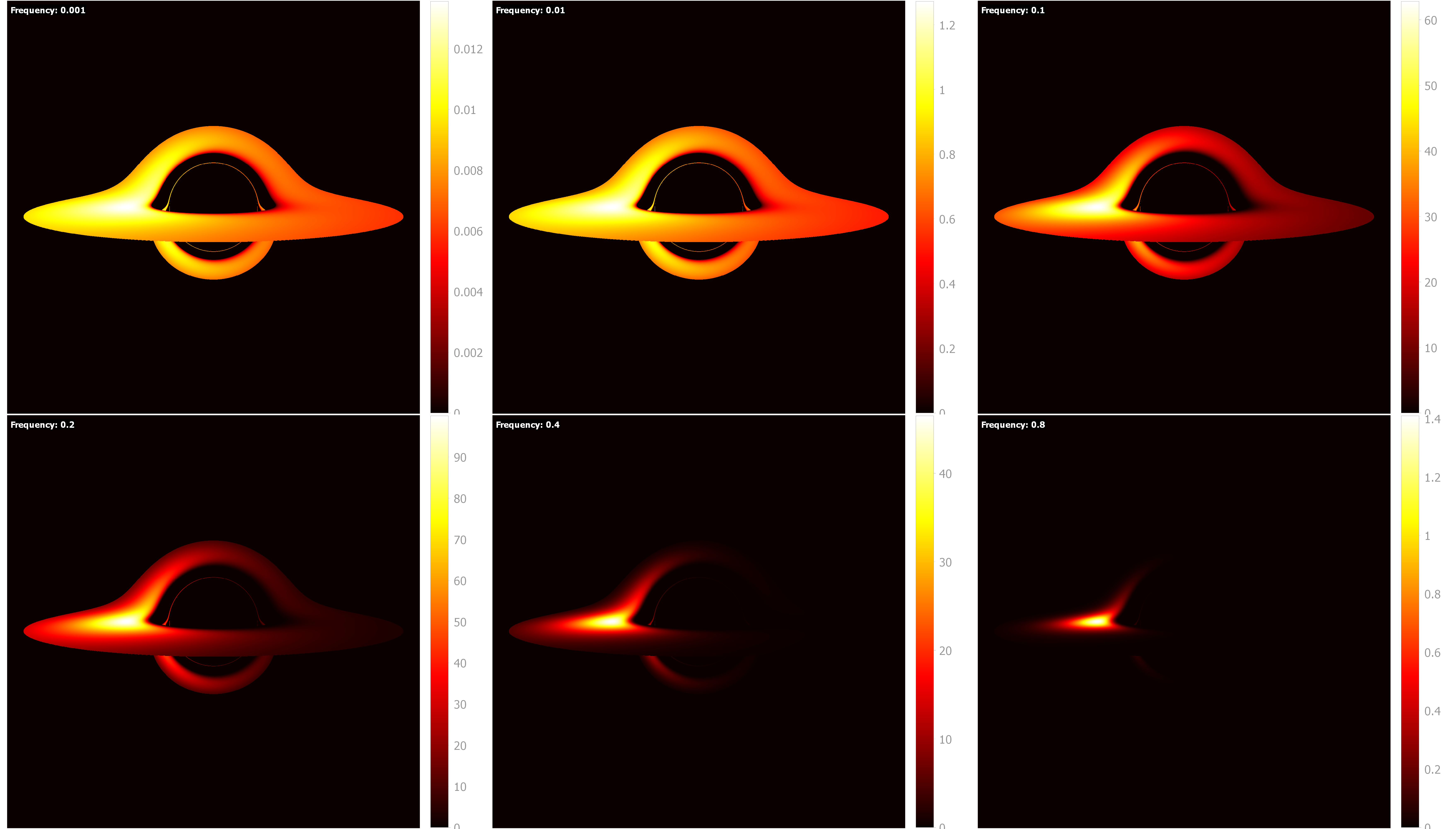}}\label{SI_0_a}\\
  \subfloat[][$\bar{\theta}=45^\circ$]{
  \includegraphics[scale=0.068]{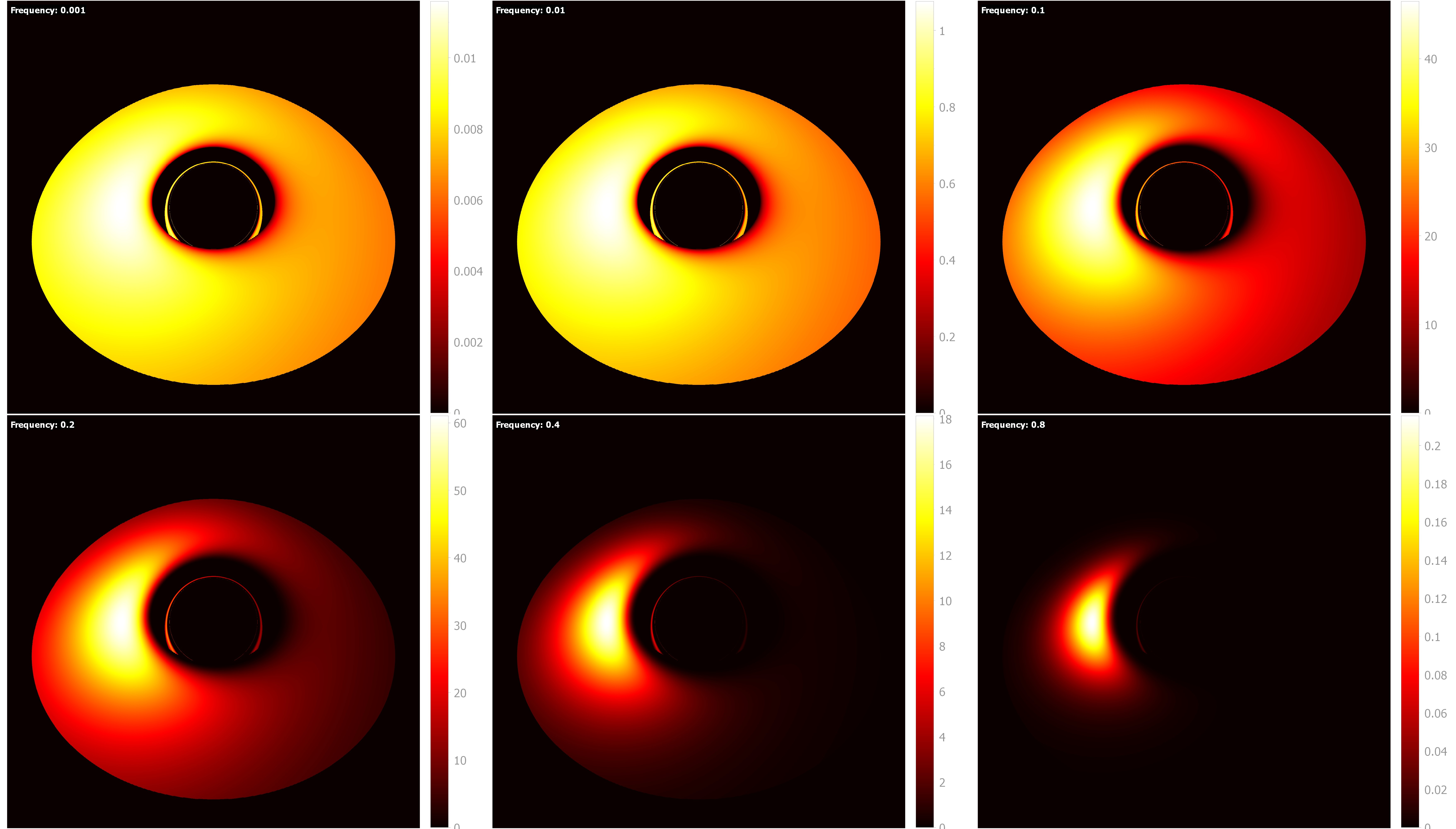}}\label{SI_0_b}
\caption{Observed specific intensity $  I(\omega_o)/I^{Sch}_{max},\textbf{\%}$ for vacuum $\omega^2_p=0$ at different frequencies $\omega_o$, mass accretion rate $\dot{M}=0.1$, ADM mass $M=1$ and inclination angle $\bar{\theta}=[84^\circ,45^\circ]$.}
\label{SI_0}
\end{figure}

We focus on two principal characteristics of the system. First, we examine the frequency-dependent specific intensity $I(\omega_o)$ to investigate how plasma dispersion modifies spectral line profiles. This analysis is particularly relevant for actual observational campaigns at specific frequencies. Second, to understand the impact of varying accretion rates, we compute the intensity $\mathcal{I}$ across different mass accretion rates. Different accretion rates produce distinct spectral line blending patterns, enabling efficient parameter space exploration through compact visual representations. As previously established, this methodology is uniquely applicable to plasma-mediated radiation transfer.

\begin{figure}[tb!]
\centering
 \includegraphics[scale=0.068]{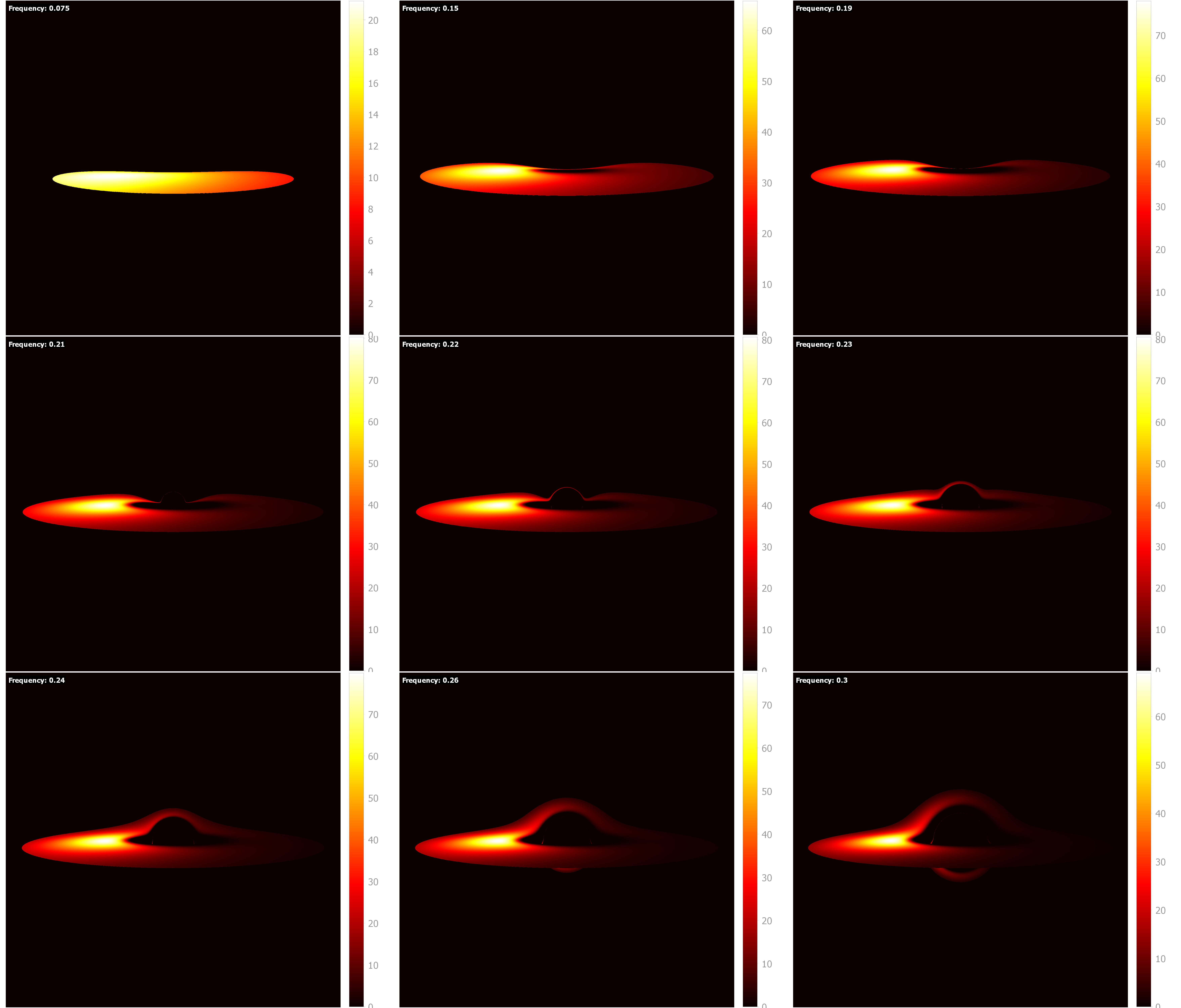}
\caption{Observed specific intensity $  I(\omega_o)/I^{Sch}_{max},\textbf{\%}$ for plasma profile $\omega^2_p=M^2 r^{-2}$ at different frequencies $\omega_o$, mass accretion rate $\dot{M}=0.1$, ADM mass $M=1$ and inclination angle $\bar{\theta}=84^\circ$.}
\label{SI_2_a}
\end{figure}

\begin{figure}[tb!]
\centering
 \includegraphics[scale=0.068]{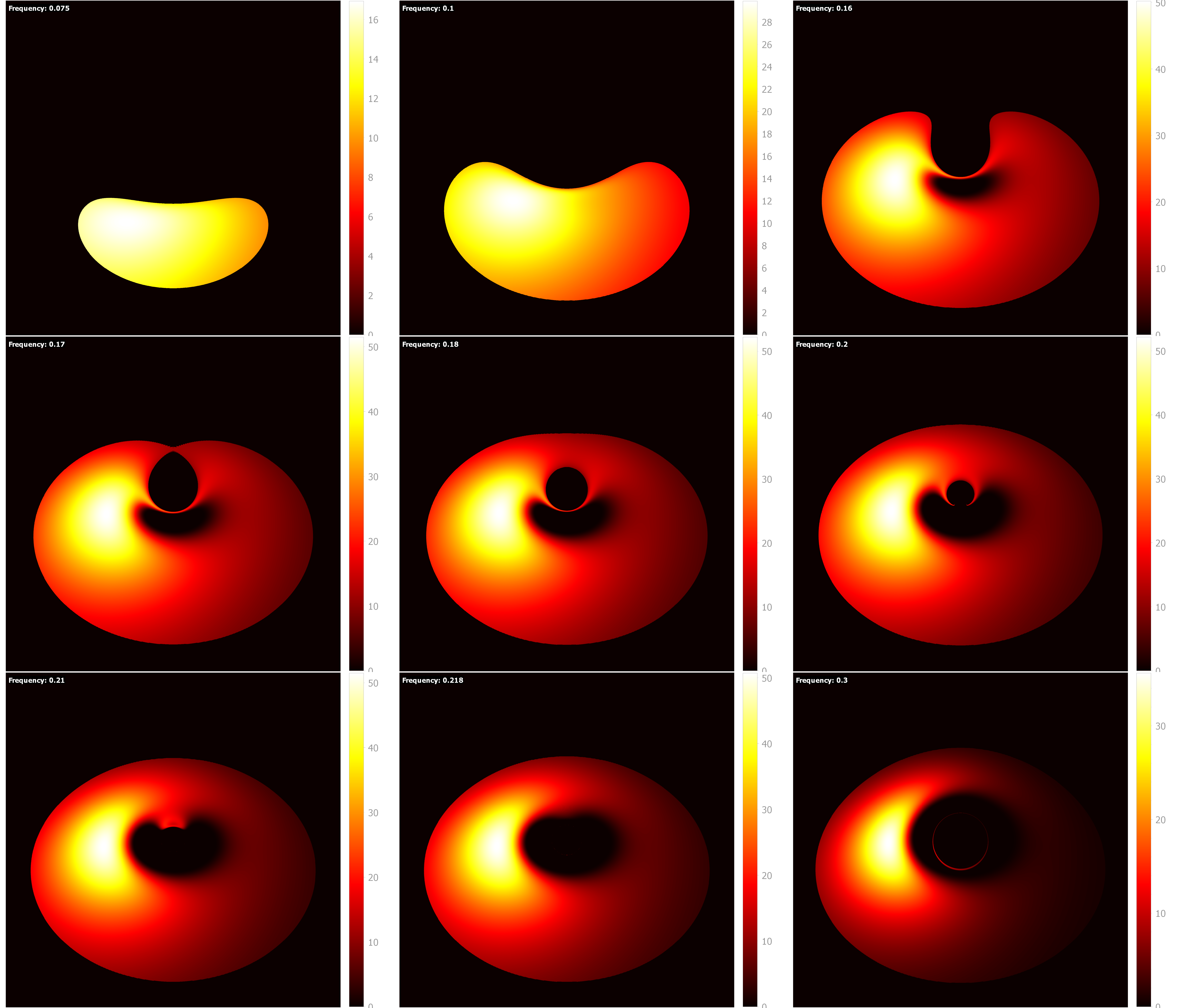}
\caption{Observed specific intensity $  I(\omega_o)/I^{Sch}_{max},\textbf{\%}$ for plasma profile $\omega^2_p=M^2 r^{-2}$ at different frequencies $\omega_o$, mass accretion rate $\dot{M}=0.1$, ADM mass $M=1$ and inclination angle $\bar{\theta}=45^\circ$.}
\label{SI_2_b}
\end{figure}

Figure~\ref{SI_0} displays characteristic radiation intensity distributions from a thin accretion disk in a vacuum observed at the inclination angles $\bar{\theta} = [84^\circ,45^\circ]$, for various frequencies at a fixed mass accretion rate $\dot{M} = 0.1$ and $M=1$. The intensity is normalized to the maximum radiation intensity in the vacuum Schwarzschild metric at the inclination angle $\bar{\theta} = 84^\circ$ and represented as percentages on the color scale. For visualization purposes, each color scale is independently normalized to the maximum intensity at its respective frequency - for instance, the maximum intensity for $\omega_o=0.01$ is notably lower than for $\omega_o=0.2$.

The intensity distribution exhibits strong frequency dependence, reflecting the unique temperature profile across the disk. Lower frequencies $\omega_o$ produce smoother intensity gradients, while higher frequencies yield sharper transitions. This behavior originates from the inherent asymmetry of the Planck spectrum and persists independently of dispersion effects, as will be evident in all subsequent results.

Let us examine in detail the plasma distribution with $\omega^2_p = M^2/r^2$, whose corresponding accretion disk images are shown in Figs.~\ref{SI_2_a} and ~\ref{SI_2_b}. The equilibrium frequency for this configuration, calculated via Eq.~(\ref{eq:equilibrium_f}), is $\omega_e = 0.19745$. For frequencies below equilibrium, Figs. \ref{SI_2_a}-\ref{SI_2_b} demonstrates the complete absence of both gravitational shadows and relativistic images \cite{Virbhadra:1999nm,Virbhadra:2008ws} due to strong plasma reflection. In essence, only the primary image of a front edge of the accretion disk is observed. In subsequent images $\omega_o=[0.2-0.3]$, the gravitational shadow emerges, surrounded by emission from the disk's distant region. However, for inclination angle $\bar{\theta}=84^\circ$ the image is still much flatter than in a vacuum Fig. \ref{SI_2_a}. The peculiarity of inclination angle $\bar{\theta}=45^\circ$ is the large area of additional disk images Fig. \ref{SI_2_b}. At the same time, this configuration produces a striking visual effect resembling binary system shadows. As we will see later, this will also be clearly noticeable when considering the total observed radiation flux. As the frequency increases $\omega_o\geq0.3$, relativistic images become gradually distinguishable. At higher frequencies, the differences from the plasma-free case become minimal. 

\begin{figure}[tb!]
\centering
\includegraphics[scale=0.068]{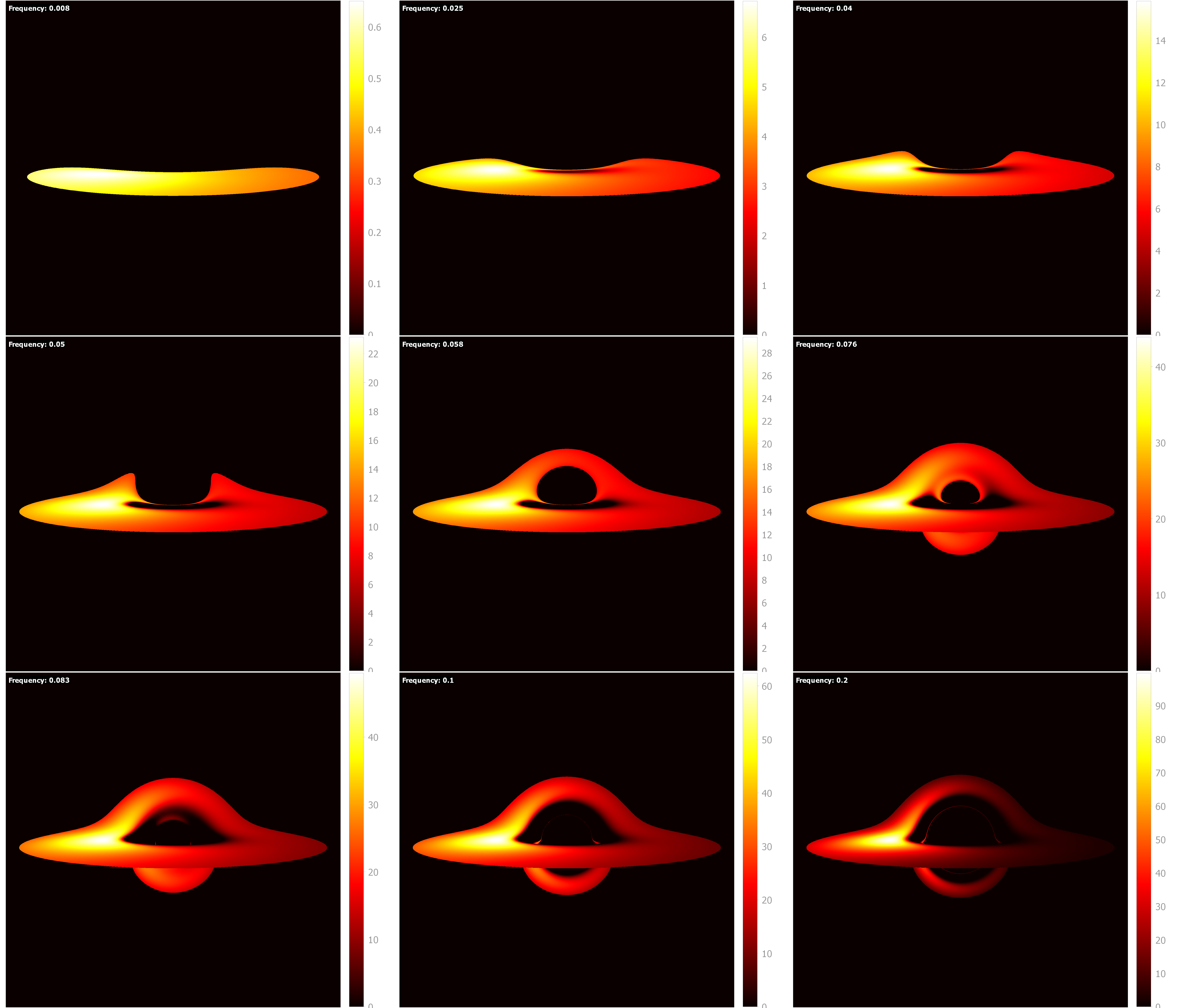} 
\caption{Observed specific intensity $  I(\omega_o)/I^{Sch}_{max},\textbf{\%}$ for plasma profile $\omega^2_p=M^4r^{-4}$ at different frequencies $\omega_o$, mass accretion rate $\dot{M}=0.1$, ADM mass $M=1$ and inclination angle $\bar{\theta}=84^\circ$.}
\label{SI_4_a}
\end{figure}

\begin{figure}[tb!]
\centering
\includegraphics[scale=0.068]{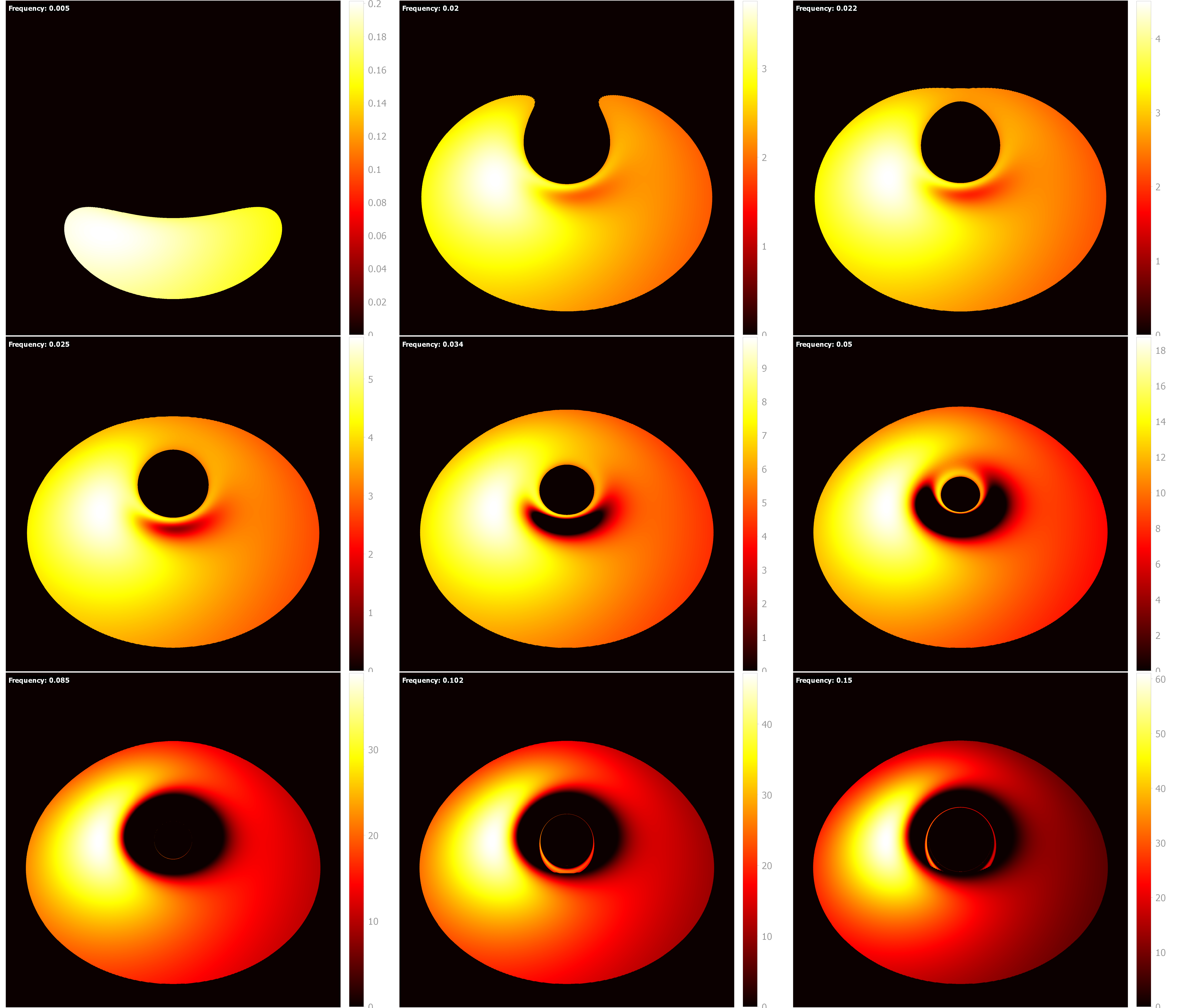} 
\caption{Observed specific intensity $  I(\omega_o)/I^{Sch}_{max},\textbf{\%}$ for plasma profile $\omega^2_p=M^4r^{-4}$ at different frequencies $\omega_o$, mass accretion rate $\dot{M}=0.1$, ADM mass $M=1$ and inclination angle $\bar{\theta}=45^\circ$.}
\label{SI_4_b}
\end{figure}

The example with plasma distribution $\omega^2_p = M^4/r^4$, presented in Figs.~\ref{SI_4_a} and ~\ref{SI_4_b}. In particular, Figs.~\ref{SI_4_a} reveals more complex image modifications compared to previous cases. At the lowest frequency $\omega_o=0.008$, the observed intensity morphology resembles that of the earlier plasma configuration. However, near the equilibrium frequency $\omega_e = 0.07341$, i.e. $\omega_o=[0.04-0.083]$, an additional secondary disk image \cite{Virbhadra:1999nm,Virbhadra:2008ws} emerges - distinct from standard relativistic images which gradually merge with the shadow. This differs from the plasma-free case, where such features typically concentrate near the gravitational shadow boundary. The secondary images of the disk front edge are clearly distinguishable and can be possibly observed in the experiment. At the same time, at frequencies $\omega_o=[0.083-0.2]$, a compressed gravitational shadow is clearly visible, which is consistent with the analytical result (\ref{eq:shadow_sch}). For inclination angle $\bar{\theta}=45^\circ$ the picture Fig. \ref{SI_4_b} is closer to the previous one Fig. \ref{SI_2_b}, however the main features appear at other frequencies, in particular a noticeable shift in the peak intensity of the radiation flux can be expected.

\begin{figure}[tb!]
\centering
  \includegraphics[scale=0.068]{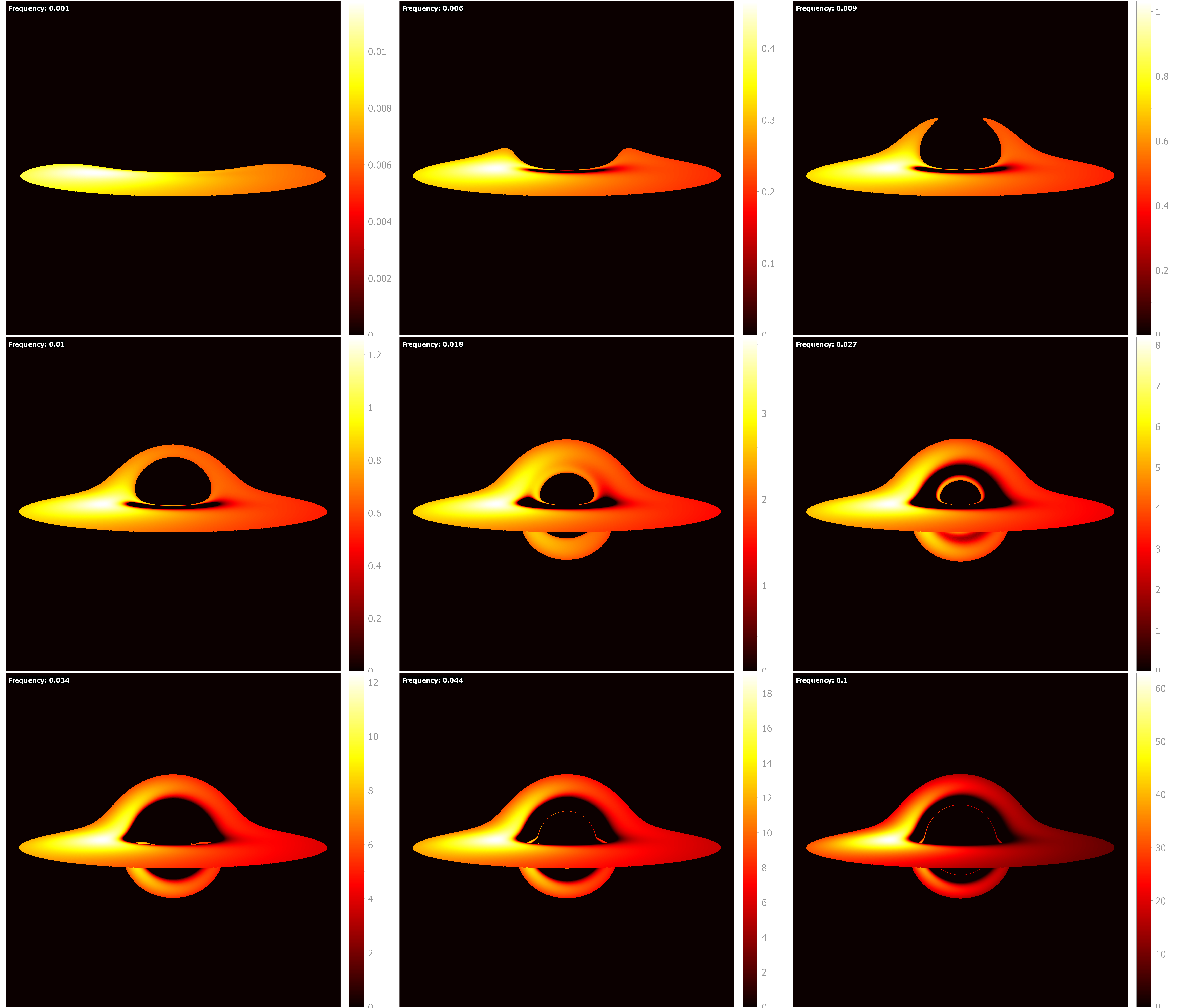} 
\caption{Observed specific intensity $  I(\omega_o)/I^{Sch}_{max},\textbf{\%}$ for plasma profile $\omega^2_p=M^6r^{-6}$ at different frequencies $\omega_o$, mass accretion rate $\dot{M}=0.1$, ADM mass $M=1$ and inclination angle $\bar{\theta}=84^\circ$.}
\label{SI_6_a}
\end{figure}

\begin{figure}[tb!]
\centering
  \includegraphics[scale=0.068]{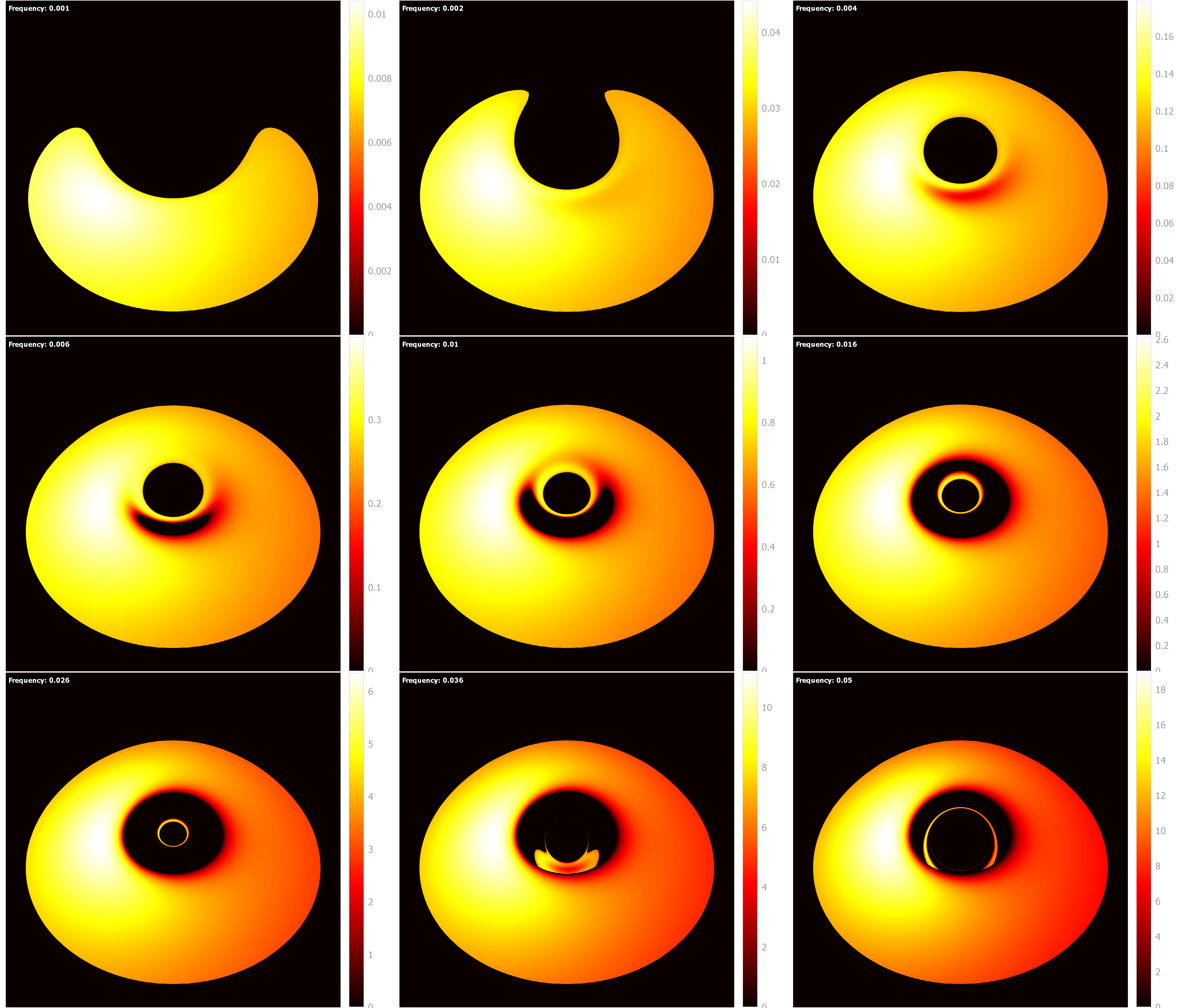} 
\caption{Observed specific intensity $  I(\omega_o)/I^{Sch}_{max},\textbf{\%}$ for plasma profile $\omega^2_p=M^6r^{-6}$ at different frequencies $\omega_o$, mass accretion rate $\dot{M}=0.1$, ADM mass $M=1$ and inclination angle $\bar{\theta}=45^\circ$.}
\label{SI_6_b}
\end{figure}

An example of plasma with a plasma frequency of $\omega^2_p=M^6/r^6$ generally exhibits similar behavior Fig. \ref{SI_6_a} and \ref{SI_6_b}, but more pronounced in the vicinity of the equilibrium frequency $\omega_e=0.0305$. A major role in this is also played by secondary images of the disk near edge, which are usually not easy to observe in the absence of plasma. This feature arises because the plasma density increases very quickly near the horizon, but has almost no effect on distant regions, as a result of which the strongest effect is on secondary and relativistic images of the disk. Thus, a specific characteristic of plasma with a rapidly increasing density is the observability of secondary images of the accretion disk radiation, while for a slowly increasing density cause  deformation of the primary images and the absorption of secondary ones. 

\begin{figure}[tb!]
\centering
\includegraphics[scale=0.048]{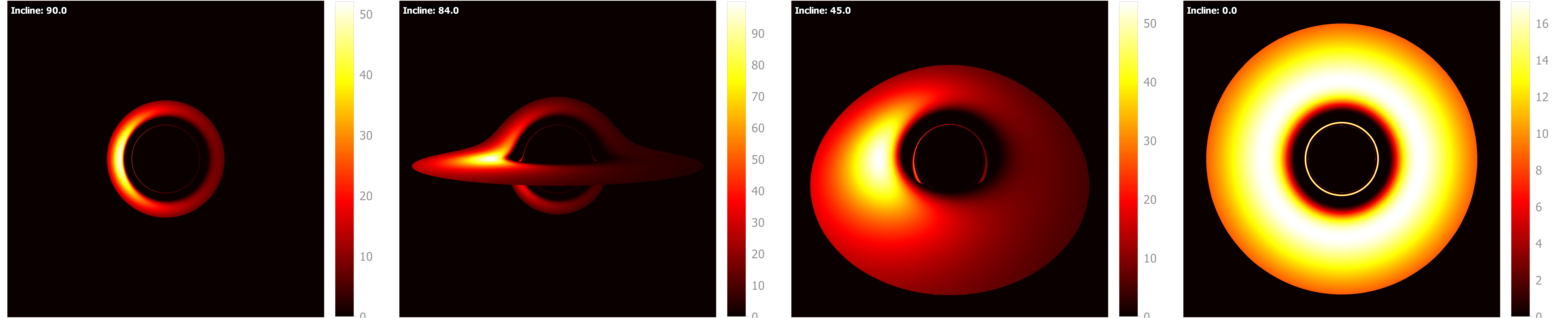} 
\caption{Observed intensity $  \I/\I^{Sch}_{max},\textbf{\%}$ for vacuum $\omega^2_p=0$, mass accretion rate $\dot{M}=0.1$, ADM mass $M=1$ and inclination angles $\bar{\theta}=[90^\circ,84^\circ,45^\circ,0^\circ]$.}
\label{I_none}
\end{figure}

As previously discussed, the radiation intensity distribution exhibits frequency dependence even in plasma-free scenarios. However, the normalized total intensity $\mathcal{I}_{\text{norm}}$ represents a particularly valuable quantity for plasma studies, as it possesses two key properties:
\begin{enumerate}
    \item Frequency independence (being an integral characteristic)
    \item Accretion-rate independence (in vacuum conditions)
\end{enumerate}

Fig.~\ref{I_none} displays this normalized total intensity distribution for the plasma-free case. Through careful normalization - verified both numerically and analytically - we confirm the distribution remains identical across all accretion rates, providing a robust baseline for plasma effect analysis.

\begin{figure}[tb!]
\centering
  \includegraphics[scale=0.2]{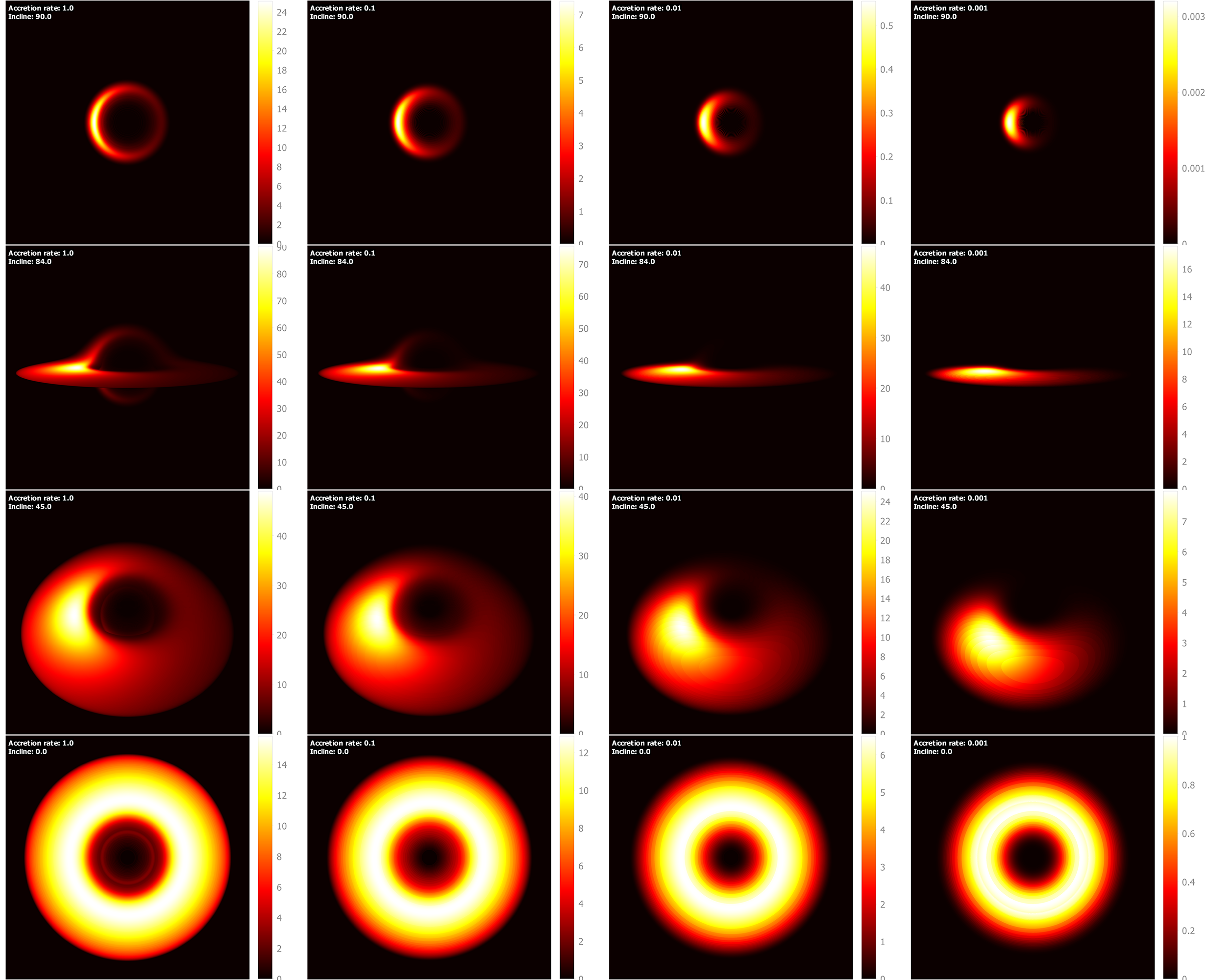} 
\caption{Observed intensity $  \I/\I^{Sch}_{max},\textbf{\%}$ for plasma profile $\omega^2_p=M^2r^{-2}$, mass accretion rates $\dot{M}=[1.0,0.1,0.01,0.001]$, ADM mass $M=1$ and inclination angles $\bar{\theta}=[90^\circ,84^\circ,45^\circ,0^\circ]$.}
\label{I_2}
\end{figure}

\begin{figure}[tb!]
\centering
  \includegraphics[scale=0.2]{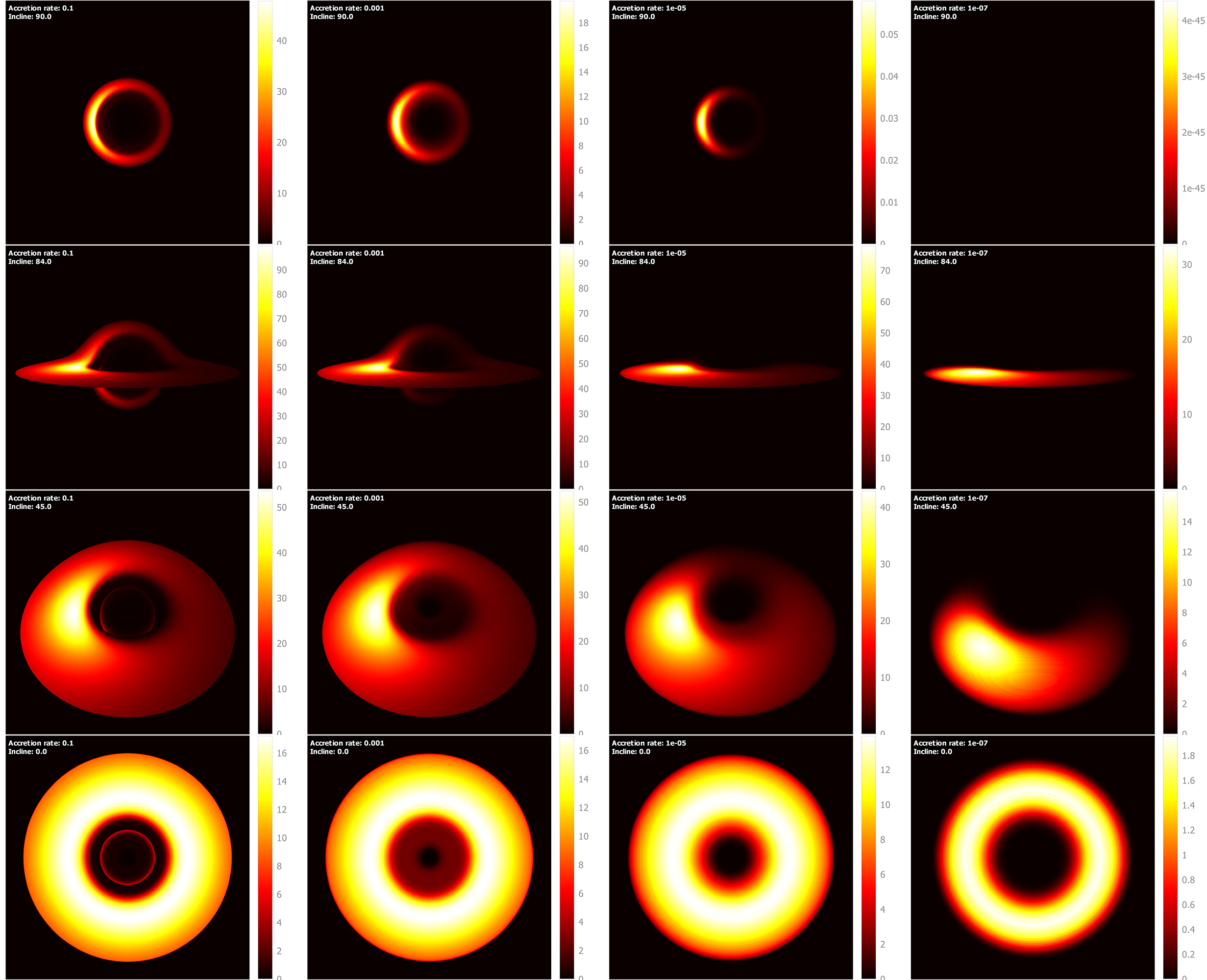} 
\caption{Observed intensity $  \I/\I^{Sch}_{max},\textbf{\%}$ for plasma profile $\omega^2_p=M^4r^{-4}$, mass accretion rates $\dot{M}=[0.1,0.001,10^{-5},10^{-7}]$, ADM mass $M=1$ and inclination angles $\bar{\theta}=[90^\circ,84^\circ,45^\circ,0^\circ]$.}
\label{I_4}
\end{figure}

\begin{figure}[tb!]
\centering
  \includegraphics[scale=0.1]{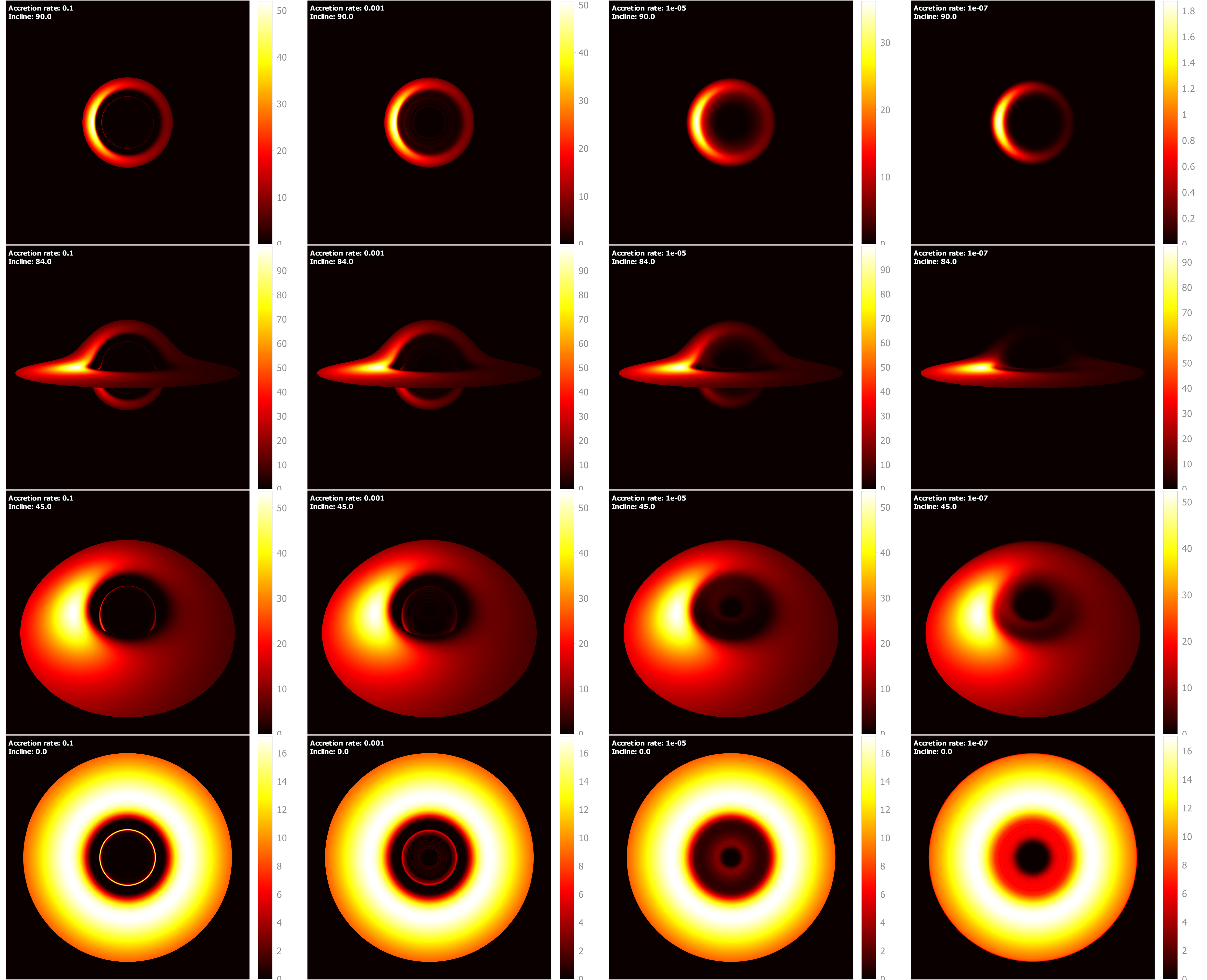} 
\caption{Observed intensity $  \I/\I^{Sch}_{max},\textbf{\%}$ for plasma profile $\omega^2_p=M^6r^{-6}$, mass accretion rates $\dot{M}=[0.1,0.001,10^{-5},10^{-7}]$, ADM mass $M=1$ and inclination angles $\bar{\theta}=[90^\circ,84^\circ,45^\circ,0^\circ]$.}
\label{I_6}
\end{figure}

When accounting for plasma effects Figs. \ref{I_2}-\ref{I_6}, we compute the total intensity (Eq.~\ref{eq:sum}) through discrete summation over frequencies:
\begin{equation}
\I \approx \sum_i I(\omega_i)\Delta \omega_i
\end{equation}
where the temperature-dependent contribution of each frequency component varies with the accretion rate. This approach reveals two significant plasma-induced phenomena:

\begin{enumerate}
    \item  Spectral line blurring occurs as different disk regions contribute to the same observed intensity through plasma dispersion effects.
    \item Systematic reduction of peak intensity emerges even without absorption processes, as evidenced in Table~\ref{table:1a_transposed}. The table quantifies this suppression as percentage decreases relative to the plasma-free case, with most pronounced effects occurring for gradual plasma distributions ($\omega_p^2 \propto r^{-\sigma}$, $\sigma \leq 2$).
\end{enumerate}

\begin{table}[h!]
    \centering
    \begin{tabular}{|c||c|c|c|c|c|c|c|c|c|}
        \hline
        \diagbox{$\omega^2_p$}{$\dot{M}$} & $10^0$ & $10^{-1}$ & $10^{-2}$ & $10^{-3}$ & $10^{-4}$ & $10^{-5}$ & $10^{-6}$ & $10^{-7}$ & $10^{-8}$ \\ \hline \hline
        $r^{-2}$ & 90 & 75 & 47 & 17 & 2 & 0 & 0 & 0 & 0 \\ \hline
        $r^{-4}$ & 99 & 99 & 98 & 96 & 90 & 77 & 56 & 32 & 13 \\ \hline
        $r^{-6}$ & 99 & 99 & 99 & 99 & 99 & 99 & 98 & 96 & 91 \\ \hline
    \end{tabular}
    \caption{Maximum intensity of observed flux $\I_{\text{max}} / \I^{\text{Sch}}_{\text{max}},\textbf{\%}$ for $\bar{\theta}=84^\circ$.}
    \label{table:1a_transposed}
\end{table}

Although we model the plasma as perfectly transparent (zero absorption), the apparent intensity reduction have careful interpretation. The observed decrease in peak intensity (Table~\ref{table:1a_transposed}) corresponds to spatial redistribution rather than energy loss. This effect manifests clearly in Figs.~\ref{I_2}-\ref{I_6}, where the conserved total energy becomes distributed over larger image areas. At the same time, some geodesics that previously contributed to the intensity do not fall on the accretion disk at all.

Unlike individual spectral slices, relativistic and secondary images are less noticeable on the total intensity map. In fact, they become very smeared due to their strong dependence on frequency, and thus do not contribute significantly to the overall intensity. For example, for mass accretion rate $10^{-7}$, the disk becomes completely invisible to an observer in the equatorial plane, as seen in Fig. \ref{I_4}. The reason for this effect is that the disk lies in the plane of observation and it is impossible to observe it through secondary images due to the strong reflection of the plasma. The case with the distribution of plasma $\omega^2_p=M^6/r^6$ is especially interesting. In particular, two dark spots in the center can be observed on the Fig. \ref{I_6}. Which may resemble the shadow of a binary system.

\begin{figure}[tb!]
\centering
  \subfloat[][]{
  \includegraphics[scale=0.5]{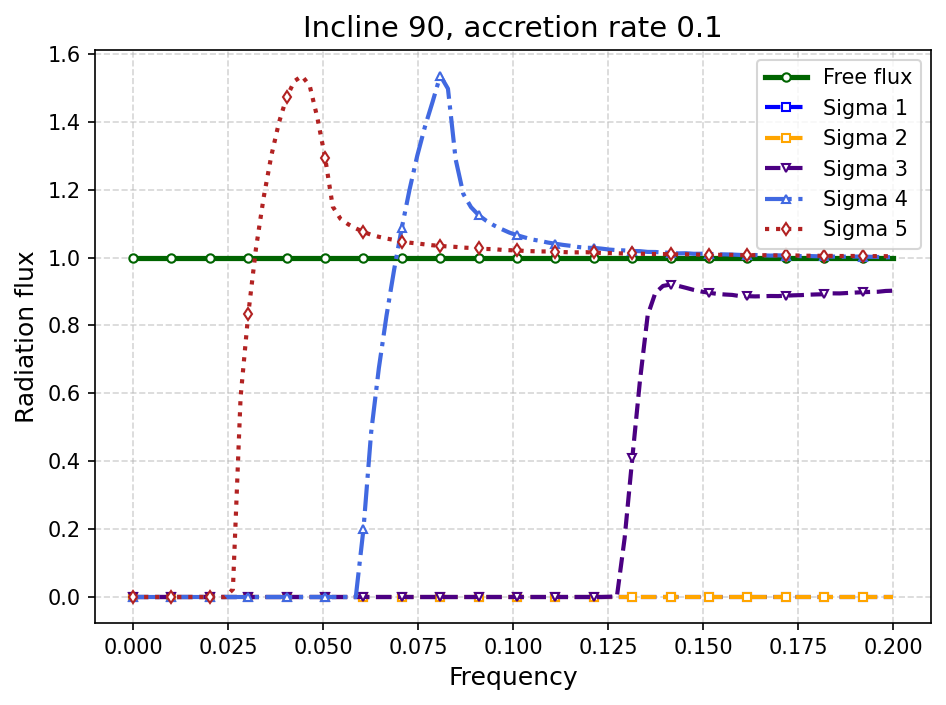} \label{flux0}
 }
  \subfloat[][]{
  \includegraphics[scale=0.5]{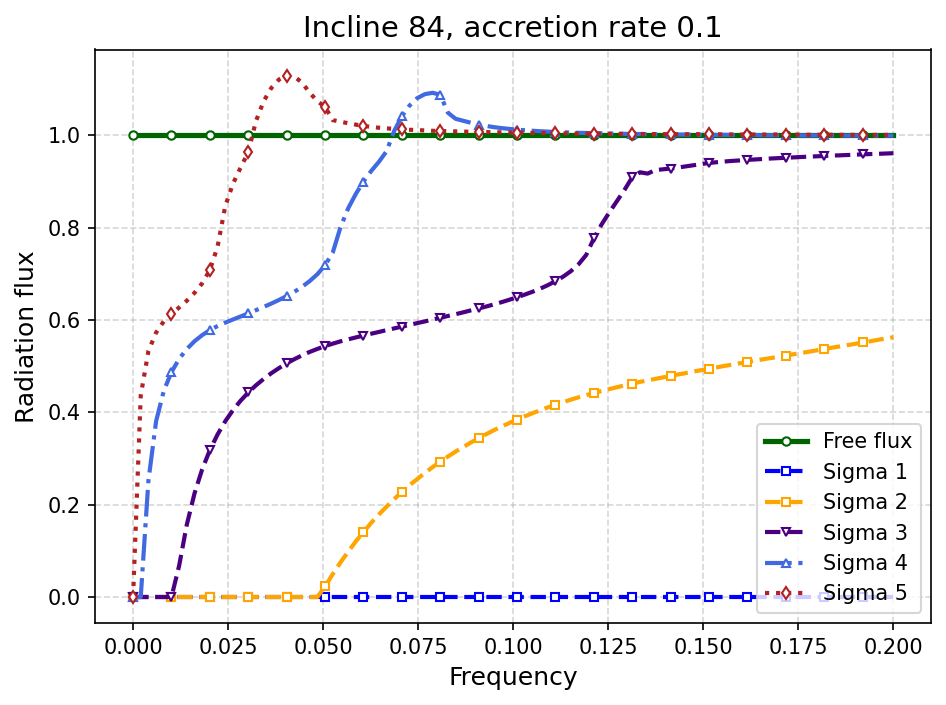} \label{flux1}
 }\\
  \subfloat[][]{
  \includegraphics[scale=0.5]{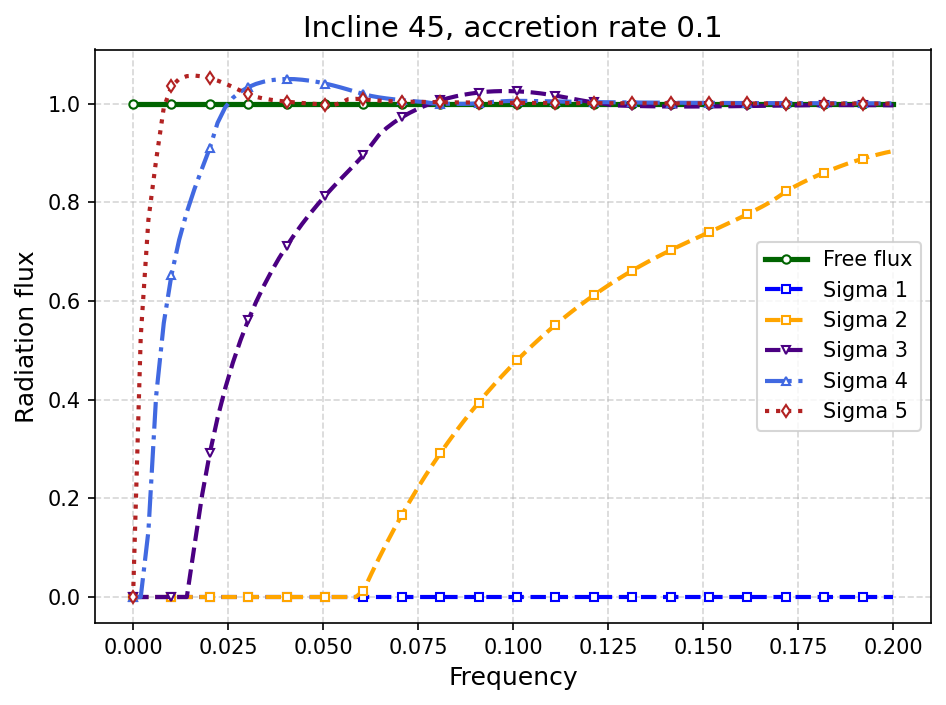} \label{flux2}
 }
   \subfloat[][]{
  \includegraphics[scale=0.5]{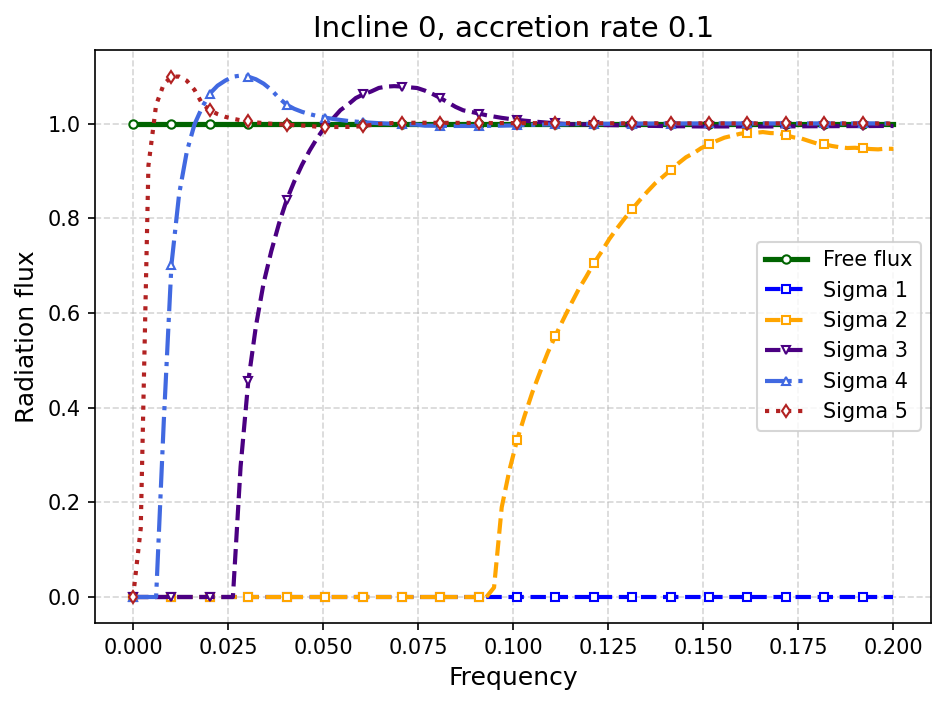} \label{flux3}
 }
\caption{Observed radiation flux $F(\omega_o)/F_{Sch}(\omega_o)$ for different inclination angles $\bar{\theta}=[90^\circ,84^\circ,45^\circ,0^\circ]$, accretion disk $r=20$ and accretion rate $\dot{M}=0.1$, ADM mass $M=1$.}
\label{flux_0}
\end{figure}

\begin{figure}[tb!]
\centering
  \subfloat[][]{
  \includegraphics[scale=0.5]{flux1.png} \label{flux1}
 }
  \subfloat[][]{
  \includegraphics[scale=0.5]{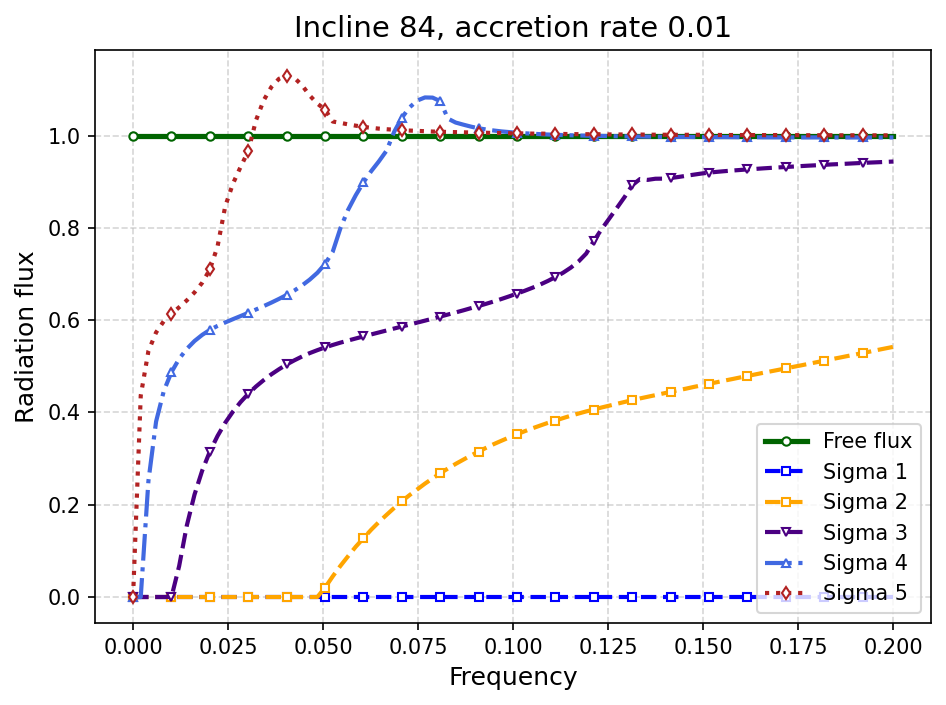} \label{flux1a}
 }\\
  \subfloat[][]{
  \includegraphics[scale=0.5]{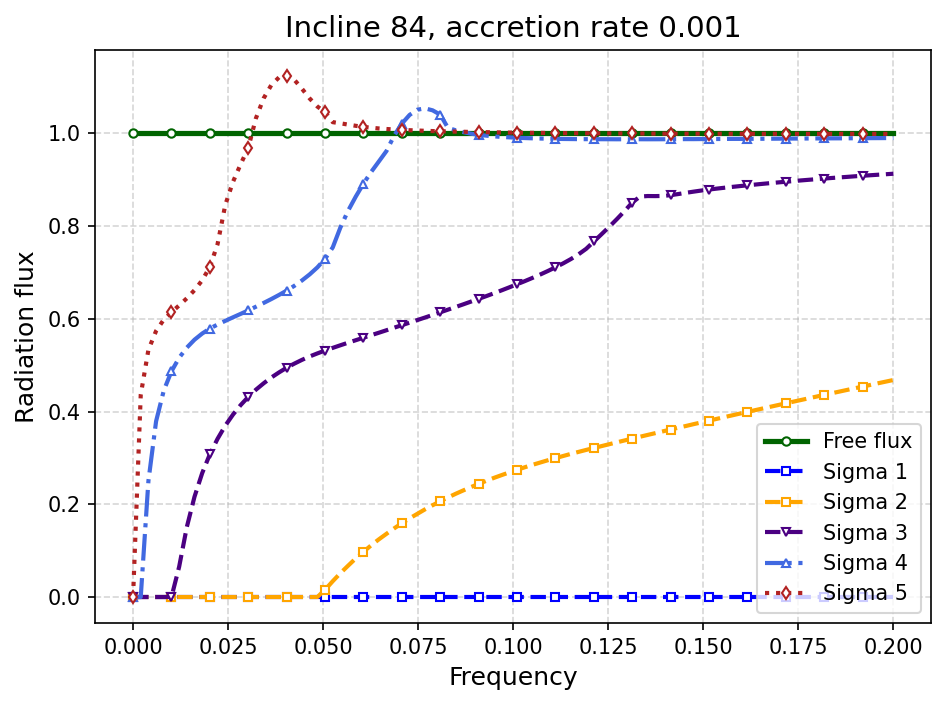} \label{flux1b}
 }
   \subfloat[][]{
  \includegraphics[scale=0.5]{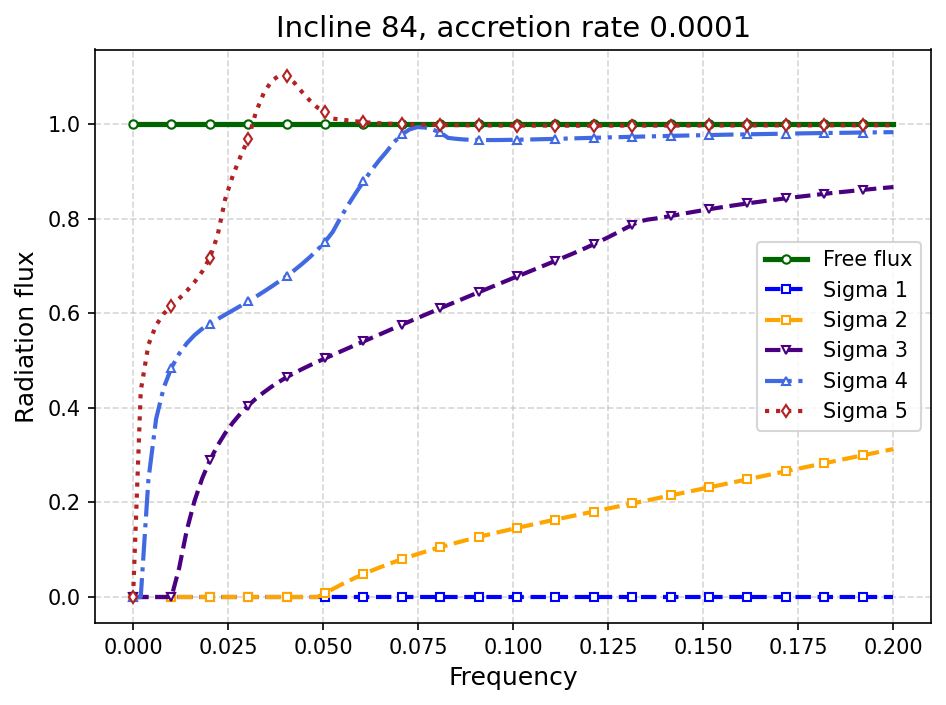} \label{flux1c}
 }
\caption{Observed radiation flux $F(\omega_o)/F_{Sch}(\omega_o)$ for different accretion rates $\dot{M}=[0.1,0.01,0.001,0.0001]$, accretion disk $r=20$ and inclination angle $\bar{\theta}=84^\circ$, ADM mass $M=1$.}
\label{flux_1}
\end{figure}

\begin{figure}[tb!]
\centering
  \subfloat[][]{
  \includegraphics[scale=0.5]{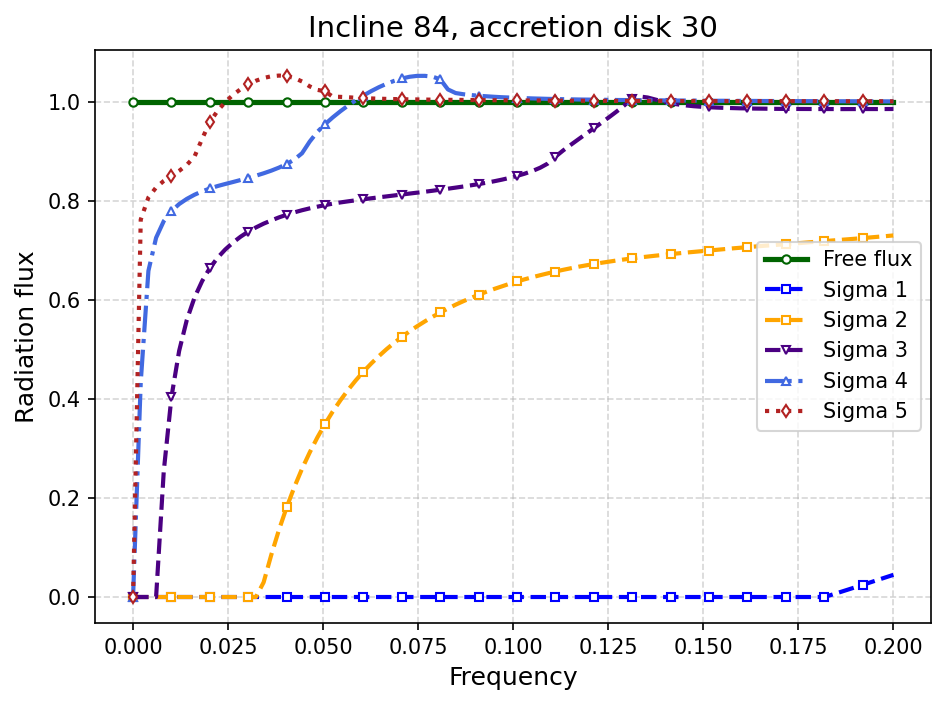} \label{flux1_}
 }
  \subfloat[][]{
  \includegraphics[scale=0.5]{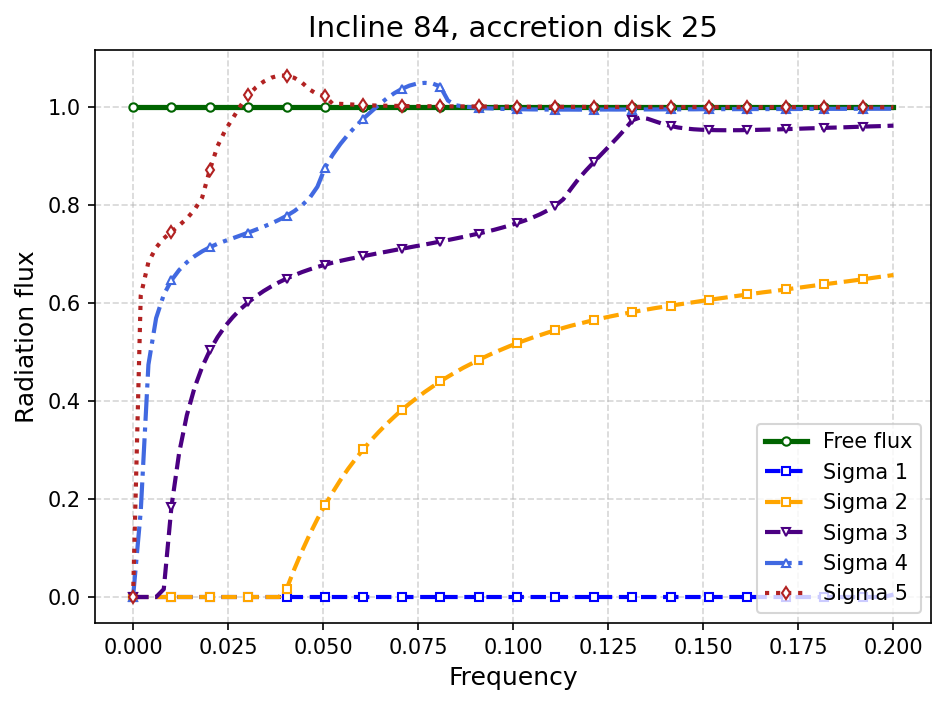} \label{flux1_a}
 }\\
  \subfloat[][]{
  \includegraphics[scale=0.5]{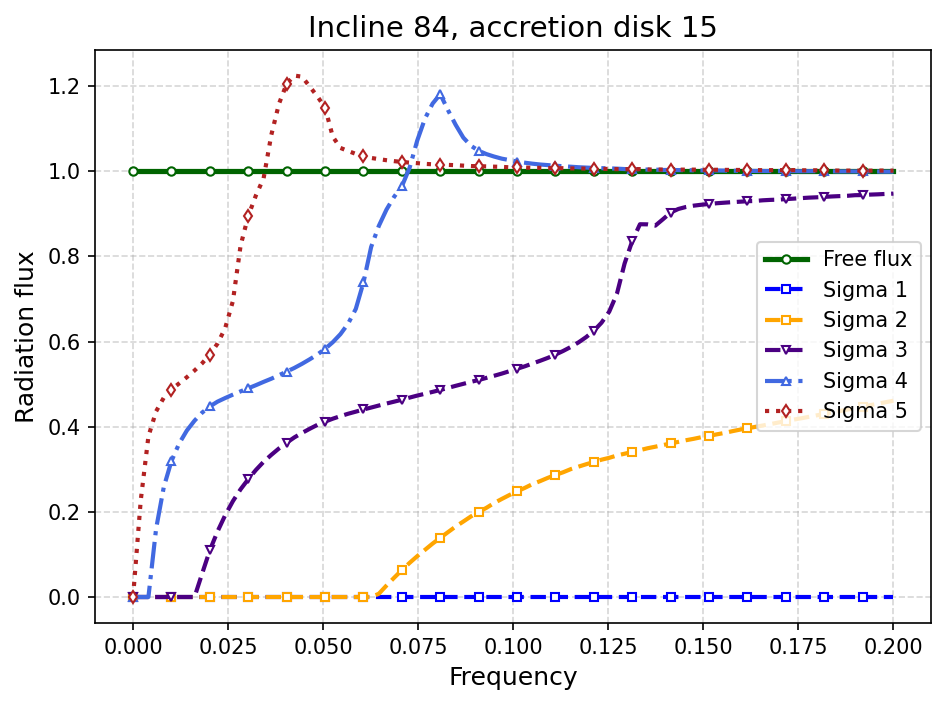} \label{flux1_b}
 }
   \subfloat[][]{
  \includegraphics[scale=0.5]{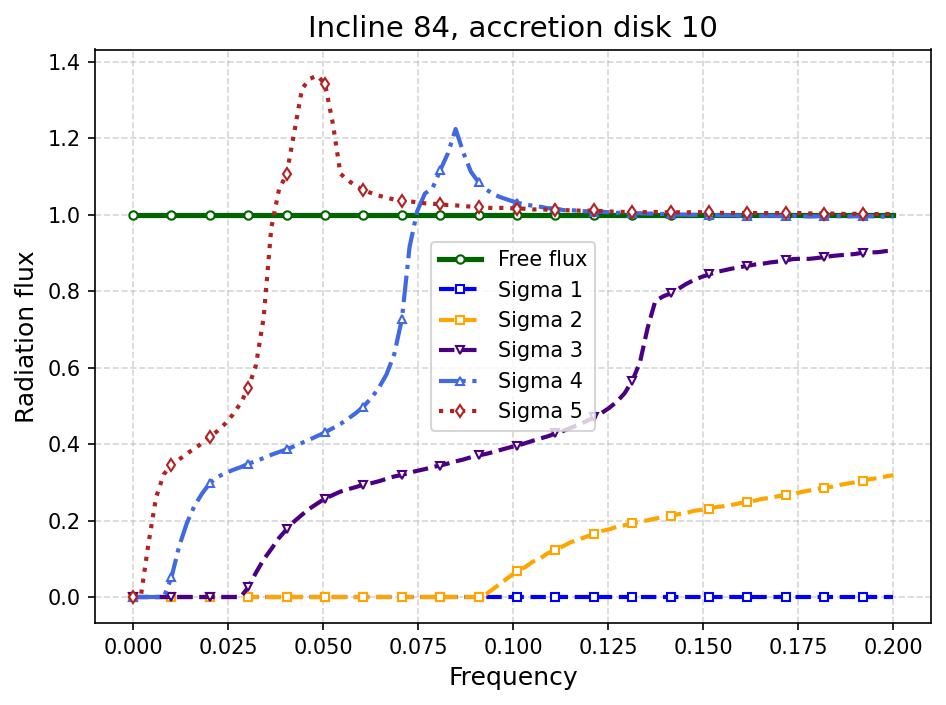} \label{flux1_c} 
  }
\caption{Observed radiation flux $F(\omega_o)/F_{Sch}(\omega_o)$ for different max radii $r=[30,25,15,10]$ of the accretion disk, inclination angle $\bar{\theta}=84^\circ$, ADM mass $M=1$.}
\label{flux_2}
\end{figure}

All these effects are quantitatively confirmed through numerical computation of the total observed radiation flux \cite{Bambi:2017khi} using the summation:
\begin{align}
F(\omega_o) = \sum_{i,j} \frac{16\cdot  I_o(\omega_o, X_{ij},Y_{ij})}{(4 +X^2_{ij}+Y^2_{ij})^2}\Delta X \Delta Y.
\end{align}
where the summation is performed over all image pixels. We examine various configurations by varying:
\begin{itemize}
    \item The inclination angle (Fig.~\ref{flux_0})
    \item The mass accretion rate (Fig.~\ref{flux_1}) 
    \item The disk size (Fig.~\ref{flux_2})
\end{itemize}
All flux values are normalized to their vacuum counterparts at the same frequency. Thus, the flux  maxima discussed throughout our analysis correspond to relative maxima.

We first observe the universal behavior across all plasma distributions: the total radiation flux approaches zero at low frequencies and asymptotically converges to the vacuum value at high frequencies. This frequency dependence arises because the plasma's repulsive effect completely obscures the accretion disk at low frequencies and plasma dispersion effects become negligible at high frequencies. The transition between these regimes reveals characteristic spectral features unique to each plasma distribution.

For plasma distributions with $\sigma \geq 4$, we observe well-defined local maxima in the relative radiation flux that persist across all tested parameters. Notably, these maxima typically exceed the corresponding vacuum flux values $F(\omega_o)/F_{Sch}(\omega_o)=1$ in the Schwarzschild metric, often representing the global maximum. This enhancement arises from secondary disk images induced by plasma lensing effects, which provide additional contributions to the total observed flux. However, two important exceptions emerge:
\begin{itemize}
    \item At sufficiently low accretion rates (Fig.~\ref{flux1c}), the maximum flux may fall below the Schwarzschild vacuum value
    \item For milder density profiles ($\sigma = 2, 3$), pronounced flux maxima appear only for specific observation angles (Figs.~\ref{flux2}--\ref{flux3})
\end{itemize}

A key finding is that the frequency position of flux maxima depends mostly on the observation angle. This fundamental property is evident from direct comparison of Figs.~\ref{flux_0}--\ref{flux_2}, where varying inclination angles (Fig.~\ref{flux_0}) systematically shift the peak frequencies while altered accretion rates (Fig.~\ref{flux_1}) and disk sizes (Fig.~\ref{flux_2}) shift slightly spectral peak locations. This angular dependence provides a direct observational diagnostic: measuring the flux maximum frequency yields the inclination angle.

While variations in the accretion rate leave the spectral positions of flux maxima almost unchanged, they significantly affect their amplitude. As shown in Fig.~\ref{flux_1}, reducing $\dot{M}$ systematically decreases the peak heights relative to the vacuum flux. This suppression becomes particularly pronounced at low accretion rates, where only high-$\sigma$ plasma distributions ($\sigma \geq 4$) maintain flux maxima exceeding their vacuum counterparts (Fig.~\ref{flux1c}). The persistence of these enhanced peaks underscores the robustness of high-$\sigma$ plasma lensing effects even in low-luminosity regimes.

Changing the size of the accretion disk also has a noticeable effect on the total radiation flux Fig. \ref{flux_2}. For small accretion disks, since most of the disk is immersed in high plasma densities, characteristic intensity peaks become markedly enhanced. The same situation occurs for an equatorial observer Fig. \ref{flux0}, since in the latter case only additional images of the accretion disk are observed, formed by geodesics passing close to the horizon.

\section{Conclusion} 
\label{sec:conclusion}

In this paper, we used both numerical and analytical methods to analyze the gravitational shadow and the observed emission spectra of thin accretion disks, taking into account the influence of stationary rotating cold plasma. We obtained a general formula (\ref{eq:intensity_transport}) for calculating the specific intensity, assuming that in the vicinity of the disk the plasma is transparent and moves with the same velocity as the accretion disk, and the spectrum itself is a blackbody spectrum. It is noteworthy that for cold plasma the intensity transport result itself does not formally depend on the plasma density at the emission point. However, this dependence is embedded in the redshift factor and temperature determined by ray tracing, which depends on the local plasma density.

For spherically symmetric metrics and plasma density, the photon trajectories are flat. By choosing this plane as the equatorial plane, we then move to arbitrary angles of photon emission from the accretion disk by rotating the coordinate system. This is useful both from the point of view of numerical calculations and analytical work. For example, this can be used to generalize the semi-analytical results \cite{Bisnovatyi-Kogan:2022ujt} for relativistic images of accretion disks to the case of an observer not located on the disk symmetry axis. We also showed that even for a stationary rotating plasma the condition $\omega\geq\omega_p$ for wave propagation is satisfied automatically (\ref{eq:applicability}) on the equations of motion and does not require verification in the numerical procedure.

We analyzed the analytical formula for the gravitational shadow radius (\ref{eq:shadow_asymptotic}) and, in particular, determined the equilibrium conditions (\ref{eq:equilibrium_r}) and (\ref{eq:equilibrium_omega}) under which the gravitational shadow disappears, since photons are reflected from the plasma at low frequencies \cite{Rogers:2016xcc}. We illustrated various cases of plasma density distribution in Fig. \ref{SSH} and also established a simple analytical equation (\ref{eq:sigma_2}) for $\sigma=2$.

Then, a numerical analysis of the observed emission spectra of the thin accretion disk in the Schwarzschild metric was performed for various plasma density distributions, mass accretion rates and inclination angles. We found that for the analysis of frequency-dependent effects of the plasma medium, it is useful to consider both specific intensity at individual frequencies and the integrated total intensity, since the latter, after normalization, does not depend on the accretion rate in the absence of plasma.

We have established the following frequency-dependent effects. The first is a strong influence on the observability of additional disk secondary images, which become observable already at the existing accuracy if we analyze frequencies in the vicinity of the equilibrium frequency $\omega_e$. Contrary, relativistic images are not observable for very low frequencies $\omega_o<\omega_e$ at all. It is important to note that the same effect can also occur in other gravitational models, but in the absence of plasma it will not depend on the frequency. The second is a decrease in the maximum intensity of the received radiation due to the blurring effect, which is unique only for frequency-dependent models. 

Another set of observed effects was extracted from analyzing the total radiative flux as a function of frequency (Figs. \ref{flux_0}-\ref{flux_2}). We found that at certain frequencies, clearly defined local maxima of the relative radiation flux appear (more precisely, plasma to vacuum flux ratio). In most cases, these maxima exceed the corresponding radiation from the accretion disk in the Schwarzschild vacuum metric, meaning they represent the global maximum of the radiation flux. Crucially, the inclination angle is the primary parameter controlling the frequency positions of these flux maxima, allowing their use for determining inclination angles. The size of the accretion disk also significantly affects the total radiation flux - for small disks where most emission occurs in high-density plasma regions, the flux intensity peaks become particularly pronounced and may experience slight frequency shifts. These findings highlight the importance of multi-frequency observations and detailed spectral analysis for identifying unique plasma effects and potential new physics beyond general relativity.

As further directions for the development of this work, we would like to highlight: generalization of the model to stationary spaces and cases of non-integrable equations of motion, taking into account the absorption of the environment using double ray-tracing, consideration of other types of plasma with an arbitrary refractive index and radiation into an anisotropic medium, as well as automation of the analysis of the obtained spectral distributions, in particular, using neural networks.

\begin{acknowledgments}
The work was supported by the Foundation for the Advancement of Theoretical Physics and Mathematics ’BASIS’.
\end{acknowledgments}

\bibliography{main}

\end{document}